\documentclass[aip,jcp,amsmath,amssymb,reprint,longbibliography]{revtex4-1}
\usepackage{graphicx} 
\usepackage{amssymb}
\usepackage{amsmath}
\usepackage{epstopdf}
\usepackage{xcolor}
\usepackage{xr}
\usepackage{lmodern}
\usepackage[version=3]{mhchem} 
\usepackage[colorlinks=False,citecolor=blue]{hyperref}%
\newcommand\gsz{0.97}

\begin{document}


\title{From Octopus to Dendrite - Semiflexible Polyelectrolyte Brush Condensates in Trivalent Counterion Solution} 



\author{Lei Liu}
\affiliation{Korea Institute for Advanced Study, Seoul 02455, Republic of Korea}
\author{Changbong Hyeon}
\thanks{hyeoncb@kias.re.kr}
\affiliation{Korea Institute for Advanced Study, Seoul 02455, Republic of Korea}

\date{\today}

\begin{abstract}
Interplay between counterion-mediated interaction and stiffness inherent to polymer chain 
can bring substantial complexity to the morphology and dynamics of polyelectrolyte brush condensates.
Trivalent counterions induce collapse of flexible polyelectrolyte brushes, over a certain range of grafting density, into octopus-like surface micelles; 
however, if individual chains are rigid enough, the ion-mediated local nematic ordering assembles the brush chains into fractal-like dendritic condensates whose relaxation dynamics is significantly slower than that in the surface micelles. 
Notably, the trivalent ions condensed in the dendritic condensates are highly mobile displaying quasi-one-dimensional diffusion in parallel along the dendritic branches. 
Our findings in this study are potentially of great significance to understanding the response of cellular organization 
such as chromosomes and charged polysaccharides on membranes to the change in ionic environment. 
\end{abstract}


\maketitle 


\section{Introduction}
Responses of polyelectrolytes (PEs) to the changes in ionic environment and chain stiffness have been extensively studied in polymer sciences \cite{OosawaBook,SkolnickMacro77,Barrat93EL,ha1995macromolecules,schiessel1999macromolecules}. 
However, new discoveries on the properties of PE are still being made through studies on biopolymers \cite{Caliskan05PRL,moghaddam09JMB,liu2016BJ,emanuel2009PhysBiol}.  
Also under active investigation are the effects of other variables and constraints on the higher order organization and dynamics of PE found in biological systems \cite{needleman2004PNAS,hud2005ARBBS,2015Hyeon198102}.

Demonstrated in both experiments and computational studies \cite{2000Williams106,2006Hud8174,2001Stevens130,2001Lee3446,2005Muthukumar074905,2013Bachmann028103,2014Maritan064902}, 
even the conformational adaptation of a single PE chain can be highly complex. 
Whereas flexible PE chains form compact globules in the presence of counterions \cite{schiessel1999macromolecules}, 
the same condition drives semiflexible PE chains (e.g., dsDNA) to toroidal conformations or metastable rod-like bundles comprised of varying number of racquet structures. 
Geometrical constraints such as confinement \cite{morrison2009PRE,spakowitz2003PRL} and increasing density of PE could add another level of complexity to the system.
For example, DNA chain in a viral capsid or nuclear envelope adopts a densely packed structure with the swelling due to the volume exclusion being suppressed by the confinement and counterions \cite{2001Kenneth14925,2013Leforestier201,Berndsen14PNAS,2015Hyeon198102}. 
Further, statistically distinct conformations of DNA emerge depending on the amount and type of counterions being added \cite{2015Nordenskiold8512,yoo2016NatComm}.

PE brush \cite{1991Pincus2912,1994Zhulina3249}, 
a spatial organization with one end of many PE chains densely grafted to 2D surface, is another system of interest to be studied. 
In particular, the novel functions and adaptability discovered in biopolymer brush \cite{2007Israelachvili1693} deserve much attention. 
For example, brush layer of hyaluronic acid, a negatively charged flexible polysaccharide molecule serving as a main component of the extracellular matrix \cite{2012Richter1466}, modulates the interaction between cells and their environment \cite{2004Addadi1393}. 
The brush of Ki-67, a disordered protein bearing a large positive net charge, found on the surface of mitotic chromosomes prevents aggregation of chromosomes \cite{2016Daniel308}.

Morphology of a polymer brush condensed in poor solvent has been studied by using theories and simulations for decades 
\cite{1992Binder586,1993Murat3108,1993Williams1313,1998Zhulina1175,2005Pereira214908,2009Dobrynin13158,2010Szleifer5300,2014Sommer104911,2016Terentjev1894}. 
Depending on the chain stiffness, brush condensates adopt diverse morphological patterns that vary from semi-spherical octopus-like micelle domains to cylindrical bundles of rigid chains which protrude from the grafting surface.  
It was shown that when the grafting density is in a proper range, multivalent counterions can collapse flexible PE brush even in \emph{good} solvent into octopus-like surface micelles displaying substantial lateral heterogeneity \cite{2016Tirrell284,2017Hyeon1579,2017dePablo155}, which has recently been confirmed experimentally for polystyrene sulfonate brush condensates in \ce{Y(NO_3)_3} solution \cite{2017Tirrell1497}.

However, we note that the aforementioned studies on the formation of surface micelles from ion-induced collapse of PE brush are still at odds with the findings by Bracha \emph{et al.} \cite{2014BarZiv4945} which reported fractal-like dendrite domains as a result of multivalent counterion (\ce{Spd^3+}) induced collapse of DNA brush.
Although the previous studies on flexible PE brush \cite{2016Tirrell284,2017Hyeon1579,2017dePablo155} captured a number of essential features reported by Bracha \emph{et al.} \cite{2014BarZiv4945}, 
qualitative difference in the morphology of brush condensate still exists, thus requiring further investigation.
To our knowledge, 
PE brush condensates with dendritic morphology remain unexplored both in theory and computation.
To this end, we extended our former work \cite{2017Hyeon1579} to scrutinize the effect of semiflexibility of PE chain on the brush morphology and dynamics in trivalent counterion solution.

\begin{figure*}[t]
\centering\includegraphics[width=0.6\linewidth]{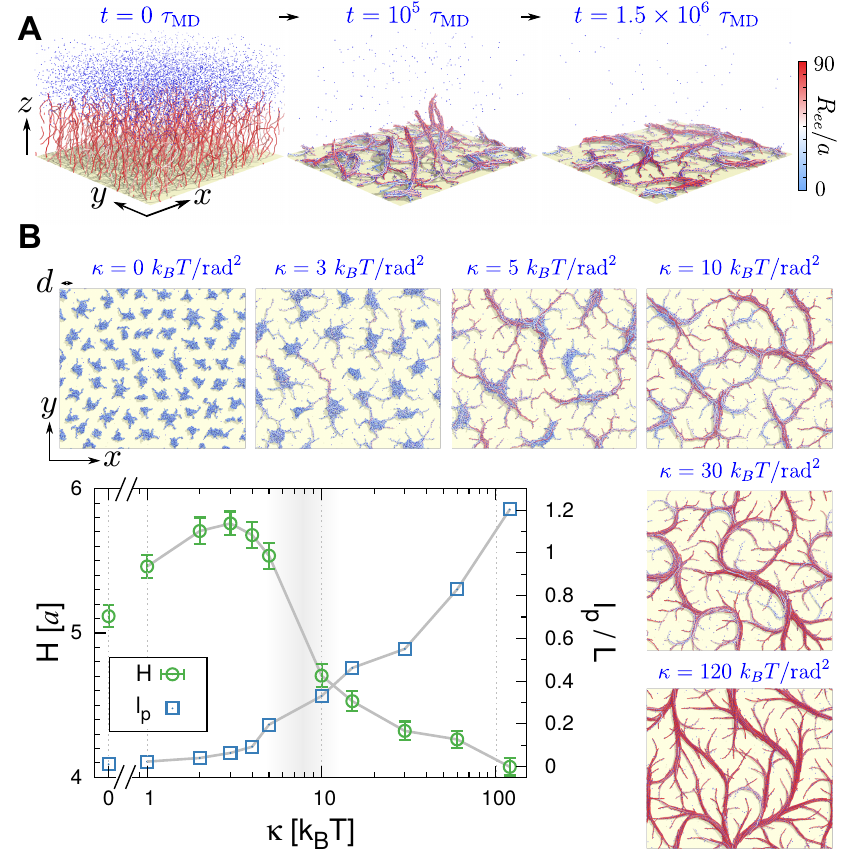}
\caption{
Model and morphologies of the brush condensates at varying chain stiffness. 
(A) The polyeletrolyte brush was modeled by $16\times16$ polymer chains, each carrying $N = 80$ negatively charged monomers, 
grafted in a triangular lattice of spacings $d=16 a$ on the surface at $z=0$. 
In the presence of trivalent cations (blue dots), 
a pre-equilibrated brush forms mutiple bundles, and eventually fully collapses onto the surface. 
For the sake of visual clarity, monovalent cations releasd from the chains, as well as monovalent anions added with trivalent cations, are not shown. 
Individual chains are color-coded from blue to red based on their end-to-end distance $R_{ee}$.
(B) Brush height $H$ and apparent persistence length $l_{p}$ of chains in the brush (see {\bf Model and Methods}) normalized by $L(=Na)$ at different $\kappa$ (main panel). 
Six snapshots of brush condensates obtained from simulations performed with different $\kappa$ are depicted in the smaller panels.
}
\label{cfg}
\end{figure*} 

In this study, we adapted a well tested coarse-grained model of strong polyelectrolyte PE brush \cite{2000Seidel2728,2003Stevens3855,2014Hsiao2900,2017Hyeon1579}. 
As shown in Fig.~\ref{cfg}A, total $M (= 16\times16)$ PE chains were grafted to the triangular lattice on uncharged surface. 
Each PE chain consists of $N (= 80)$ negatively charged monomers and a neutral terminal monomer grafted to the surface.
The lattice spacing was selected to ensure the lateral overlap between neighboring chains.  
The rigidity of PE chains was adjusted by varying the bending rigidity parameter $\kappa$.
We added trivalent salts to the pre-equlibrated PE brush in salt-free condition, and induced the collapse into brush condensate. 
Details of the model and simulation methods are given in {\bf Model and Methods}.
The results of this work are organized such that we first address the overall morphology of brush condensate under 1:3 stoichiometric condition of trivalent counterion with respect to a monovalent charge on each monomer. 
Next, the local structure of brush chain is characterized by exploiting the liquid crystal order parameters. 
Finally, we investigate the dynamics of brush condensates and of condensed counterion at varying $\kappa$ by calculating the intermediate scattering function. 

\section{Results}
{\bf Morphology of brush condensates. }
Regardless of the value of $\kappa$, the PE brush fully collapses onto the grafting surface due to the osmotic pressure of ions, which differs from neutral semiflexible polymer brushes or salt-free PE brushes in poor solvent where the aggregated bundles protrude out of the grafting surface \cite{2013Zippelius042601,2009Dobrynin13158,2014Sommer104911}. 
The morphology of the condensate depends critically on $\kappa$ (Fig.~\ref{cfg}B).
(i) For small $\kappa$ ($\alt 3$ $k_{B}T/\text{rad}^2$, $l_{p} < L/10$),
the PE brush forms octopus-like surface micelle domains demarcated by the chain-depleted grain boundaries. 
The average height of the brush $H$ 
increases with $\kappa$ ($\leq 3$ $k_{B}T/\text{rad}^2$). 
So does the surface area of the domain projected onto the $xy$-plane (see also Fig.~S1).
(ii) For large $\kappa$ ($> 15$ $k_{B}T/\text{rad}^2$, $l_{p} > L/2$), 
the condensed chains are organized into a dendritic assembly. 
Neighboring chains are assembled together, forming axially coaligned branches of varying thickness. 
The density of chain monomer slightly increases as the chain gets stiffer (see Fig.~S2A), 
which reduces brush height.
It is also noted that the end-to-end distance $R_{ee}$ of the collapsed polymers, color-coded from blue to red for individual chains, displays the broadest distribution at an intermediate stiffness $3 < \kappa < 15$ $k_{B}T/\text{rad}^2$, 
which indicates that the conformational ensemble displays the most heterogeneous distribution in this range of $\kappa$ (see also Fig.~S3A,B).

\begin{figure}[t]
\centering\includegraphics[width=\gsz\linewidth]{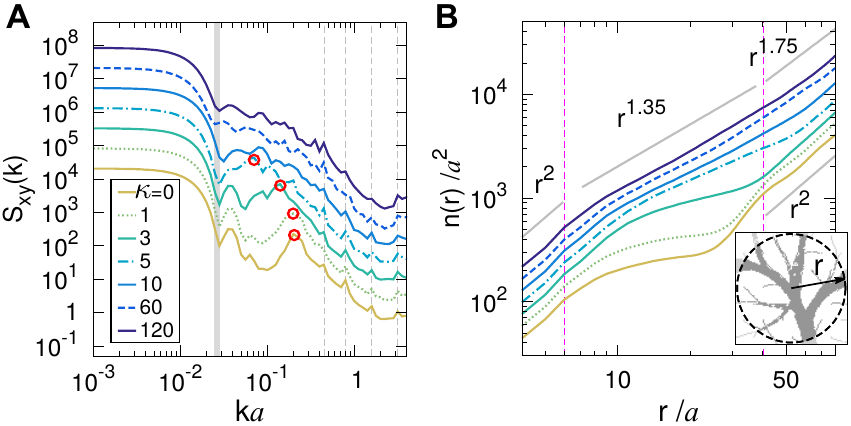}
\caption{Structure of brush condensates. 
(A) In-plane static structure factor $S_{xy}(k)$ as a function of wave number $k$ for brushes of different $\kappa$. 
The gray solid bar demarcates the range $2\pi/L_{x} \leq k \leq 2\pi/L_{y}$, i.e., the periodic boundary of the simulation box.
The red circles highlight the position of primary peak when $\kappa \leq 5$ $k_{B}T/\text{rad}^2$. 
(B) Area of the condensate on $z=0$ plane $n(r)$, as a function of a linear dimension $r$ with respect to its center. 
One exemplary illustration is provided in the inset. 
For visual clarity, the curves in (A,B) are shifted upward progressively.
}
\label{static}
\end{figure}

To quantify the in-plane lateral configuration of the brush, we calculated the 2D static structure factor 
\begin{equation}
S_{xy}(k) = \Big \langle \Big \langle \frac{1}{N_{m}} \bigg \arrowvert \sum_{i,j=1}^{N_{m}} e^{i \vec{k} \cdot (\vec{r}_i-\vec{r}_{j})} \bigg \arrowvert\Big \rangle_{|\vec{k}|} \Big \rangle,
\label{eq-ssf} 
\end{equation} 
where $N_{m} = M \times N$ is the total number of non-grafted chain monomers, 
$\vec{r}_{i}$ is the position of the $i$-th monomer, and $\vec{k}$ is a 2D wave vector in the $xy$ plane.
$S_{xy}(k)$ is evaluated by first integrating over the space of $| \vec{k} | = k$, followed by averaging over the ensemble of MD trajectories.
$S_{xy}(k)$ exhibits distinct profiles when $\kappa$ is varied (Fig.~\ref{static}A).
For octopus-like micelles, there is a primary peak (indicated by red circles) characterizing the size (area) of the domain, 
whose position shifts to a smaller wave number as $\kappa$ increases, indicating that the domain size grows with $\kappa$.
However, this peak gradually vanishes as the stiffness of chain is increased. 
The absence of the peak in $S_{xy}(k)$ is due to the morphological transition from the finite-sized surface micelles to the scale-free dendritic assembly.

To quantify the dendritic patterns in 2D, we further analyzed their fractal dimensions $\mathcal{D}_{f}$. 
We divide the grafting surface into square lattices with each cell size of $a \times a$.  
When at least one chain monomer is present in a cell, the cell contributes to the ``area" of the condensate. 
The area of dendritic pattern within a radius $r$ is $n(r) =\langle a^2 \sum_{p,q} o_{p,q} \Theta(r - r_{p,q})\rangle$, 
where $\Theta(\ldots)$ is the Heaviside step function, $o_{p,q} = \Theta[\sum_{i=1}^{N_{m}} \delta(x_{i} - pa)] \times \Theta[\sum_{i=1}^{N_{m}} \delta(y_{i} - qa)]$, 
and $r_{p,q}$ is the distance of the cell from a center of high monomer density. 
$n(r)$ was obtained by averaging over the cells with the five highest monomer density in each snapshot.  

$n(r)$ scales as $n(r) \sim r^{\mathcal{D}_f}$, and the value of the scaling exponent $\mathcal{D}_{f}$ varies at different length scale (Fig.~\ref{static}B).
(i) In the range of $6 < r/a < 40$, $\mathcal{D}_f\approx 1.35$ for brushes of rigid chains, 
whereas $\mathcal{D}_f\approx 0$ for flexible brushes. 
The transition from micelle domain with finite size ($\mathcal{D}_f \approx 0$) to scale-free dendritic assembly ($\mathcal{D}_f> 0$) is observed at $\kappa\approx 5$  $k_{B}T/\text{rad}^2$ (see Fig.~\ref{cfg}B).
(ii) At $r/a > 50$, $\mathcal{D}_f \approx 1.75$ for rigid brushes ($\kappa > 15$ $k_{B}T/\text{rad}^2$), 
and $\mathcal{D}_f\approx 2$ for flexible brushes ($\kappa \leq 5$ $k_{B}T/\text{rad}^2$).
The scaling exponent $\mathcal{D}_{f}\approx 2$ arises when the monomer is uniformly distributed on the surface such that the density of condensates $\rho_{m} = n(r)/{\pi r^2}$ is constant with respect to $r$. 
Unlike the octopus-like micelles surrounded by the chain-depleted zone, the dendritic condensate percolates over the entire surface.
Analyzing fluorescence images of dendritic condensate of dsDNA brush through a similar method \cite{2014BarZiv4945}, 
Bracha \emph{et al.} reported $\mathcal{D}_{f} = 1.75$.

Another quantity often being used to address the fractality is the 2D version of radial distribution function, 
defined as $C_{xy}(r) = \sum_{i>j} \delta(r_{i,j}-r) / \pi r^2 N_{m}$.
$C_{xy}$ scales as $r^{\mathcal{D}_{f} - 2}$ in a fractal aggregate of dimension $\mathcal{D}_{f}$ \cite{1986Sander789,2016Dossetti,2017Dossetti3523}. 
Consistent with the analysis of $n(r)$,  
$C_{xy}$ of chain monomers indeed follows a scaling $C_{xy} \sim r^{1.35 - 2} = r^{-0.65}$ in the intermediate range of $r$ when $\kappa$ is large (Fig.~S4C).  
\\

\begin{figure}[tb]
\centering\includegraphics[width=\gsz\linewidth]{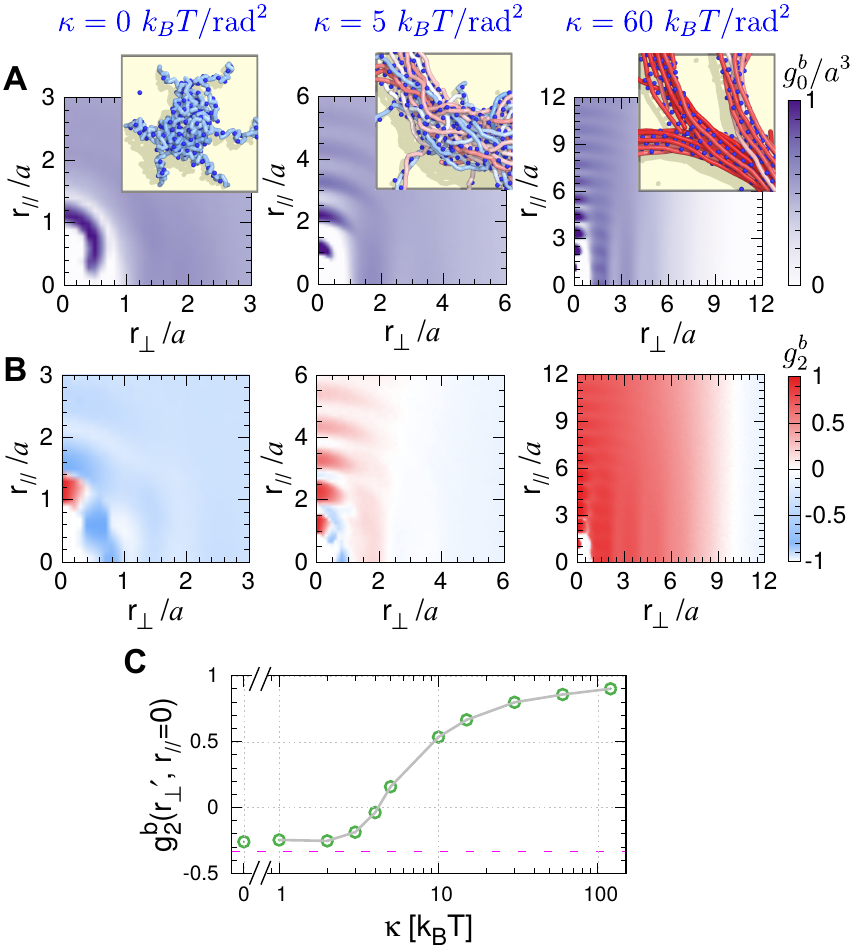}
\caption{
Local chain organizations.
(A) Heat map of the density distribution of bonds $g_{0}^{b}$ (Eq.\ref{eg0}), and (B) orientation order parameter $g_{2}^{b}$ (Eq.\ref{eg2}) as a function of $r_{\perp}$ and $r_{\parallel}$.
From left to right panels, $\kappa = 0,5,60$ $k_{B}T/\text{rad}^2$, respetively.
Insets are snapshots of a small region, with a size of $32 a\times 32 a$, in the corresponding brush condensates.
(C) Inter-chain bond orientational order $g_{2}^{b}(r_{\perp}^{\ast},0)$ as a function of $\kappa$, where $r_{\perp}^{\ast}$ is the position of the highest peak in $g_{0}^{b}(r_{\perp},0)$. 
}
\label{bdOrder}
\end{figure}

{\bf Local chain organization. }
The local structure of chains in the brush condensates also changes with $\kappa$ (see the insets of Fig.~\ref{bdOrder}A). 
When chains are flexible, the adjacent chains condensed to the same micelle appear highly entangled.
It is not visually clear whether two monomers close in space are in the same chain or in different chains. 
In contrast, when chains are rigid, they are parallelly aligned, 
and the strong orientational correlation between consecutive bonds allows us to easily discern one chain from another. 
To characterize the local ordering of polymer segments in the collapsed brush, 
we employed the liquid crystal order parameter \cite{1986Eppenga1776,2000Frenkel10034}.
For any two consecutive monomers ${i,i+1}$ in the same chain, 
a unit bond vector $\hat{b}_{i}$ is defined by its orientation $\vec{u}_{i} = (\vec{r}_{i+1} - \vec{r}_{i}) / |\vec{r}_{i+1} - \vec{r}_{i}|$, 
and its position $\vec{v}_{i} = (\vec{r}_{i+1} + \vec{r}_{i}) / 2$.
The radial distribution of such two bond vectors can be evaluated as 
\begin{equation}
g_{0}^{b}(r_{\perp},r_{\parallel}) = \frac{\sum_{i,j} \delta(r_{ij,\perp} - r_{\perp}) \delta(r_{ij,\parallel} - r_{\parallel})}{\pi r_{\perp}^2 r_{\parallel} N_{b}},
\label{eg0}
\end{equation}
where $\vec{r}_{ij}^b = \vec{v}_{j} - \vec{v}_{i}$, $\vec{r}_{ij,\parallel} = \vec{r}_{ij}^b \cdot \vec{u}_{i}^b$, 
$\vec{r}_{ij,\perp} = \vec{r}_{ij}^b - \vec{r}_{ij,\parallel}$, and $N_b = M \times N$ is the total number of bonds in the brush. 
The vector $\vec{r}_{ij}^{b}$, pointing from bond $\hat{b}_{i}$ to another bond $\hat{b}_{j}$, 
was decomposed into the parallel and perpendicular components ($\vec{r}_{ij,\parallel}$  and $\vec{r}_{ij,\perp}$ ) 
with respect to the orientation of $\hat{b}_{i}$.
The heat map of $g_{0}^{b}(r_{\perp},r_{\parallel})$ in Fig.~\ref{bdOrder}A, indicates that the bonds of flexible chains in a micelle are isotropically distributed. 
As $\kappa$ increases, density correlation first rises along the axis of $r_{\parallel}$. 
Because the effective attraction between monomers from neighboring chains increases with $\kappa$ (Fig.~S4B), 
bond density correlation also appears on the $r_{\perp}$ axis when $\kappa > 10$ $k_{B}T/\text{rad}^2$.

The relative orientational correlation between bond vectors, which cannot be described by $g_{0}^{b}(r_{\perp},r_{\parallel})$ alone, is quantified by calculating \cite{1986Eppenga1776,2000Frenkel10034}
\begin{equation}
g_{2}^{b}(r_{\perp},r_{\parallel}) = \frac {\sum_{i,j} \cos(2 \theta_{ij}) \delta(r_{ij,\perp} - r_{\perp}) \delta(r_{ij,\parallel} - r_{\parallel})} {\sum_{i,j} \delta(r_{ij,\perp} - r_{\perp}) \delta(r_{ij,\parallel} - r_{\parallel})},
\label{eg2}
\end{equation}
where $\theta_{ij}$ is the angle between $\hat{b}_{i}$ and $\hat{b}_{j}$, thus $\cos(2 \theta_{ij}) = (\vec{u}_{i} \cdot \vec{u}_{j})^2 - 1$. 
$\cos(2\theta) \leq 0$ if $\pi/4 \leq \theta \leq 3\pi/4$.
In the case of an isotropic distribution, $g_{2}^{b*} = \int_{0}^{\pi} \sin(\theta) \cos(2\theta) d\theta / \int_{0}^{\pi} \sin(\theta) d\theta = -1/3$.  
For flexible chains with $\kappa=0$ $k_{B}T/\text{rad}^2$ (Fig.~\ref{bdOrder}B left), 
the positive correlation arises only from their nearest neighboring bond along the chain, 
and $g_{2}^{b}$ converges to $-1/3$ within a very short range ($r < 2a$).
At $\kappa=5$ $k_{B}T/\text{rad}^2$, intra-chain bonds are well ordered, but on the $r_{\perp}$ axis $g_{2}^{b} \approx -1/3$ when $r_{\perp} > 2.5a$, 
which suggests that except for the nearest neighbors, the bonds from different chains are still poorly aligned.
Lastly, at $\kappa=60$ $k_BT/\text{rad}^2$, $g_{2}^{b} (r_{\perp},r_{\parallel})> 0$ in both $r_{\perp}$ and $r_{\parallel}$ directions with $r_{\perp}$, $r_{\parallel}\gg 1$, 
in agreement with the observation that rigid chains are bundled together forming the branches of the condensate.

To highlight the effect of $\kappa$ on the local \emph{inter}-chain organization in the condensate, 
we plotted $g_{2}^{b}(r_{\perp}^{\ast},0)$ (Fig.~\ref{bdOrder}C) against $\kappa$, by considering it as a single-valued estimate of the inter-chain bond alignment, 
where $r_{\perp}^{\ast}$ is position of the highest peak of $g_{0}^{b}(r_{\perp}^{\ast},0)$ (see also Fig.~S5). 
In the brush condensate, chains are randomly entangled with each other when $\kappa \leq 3$ $k_{B}T/\text{rad}^2$, 
but they display nearly perfect alignment when $\kappa > 30$ $k_{B}T/\text{rad}^2$.
This disorder-to-order ``transition" takes place around $\kappa \approx 5$ $k_{B}T/\text{rad}^2$ (Fig.~\ref{bdOrder}C).
\\

\begin{figure}[tb]
\centering\includegraphics[width=\gsz\linewidth]{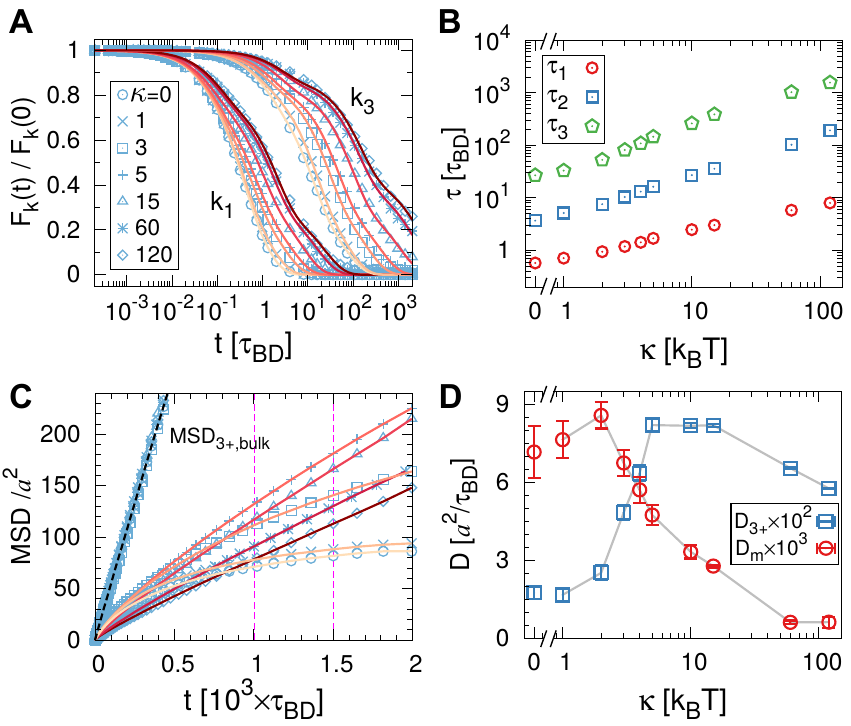}
\caption{
Dynamic properties of brush condensate and counterions. 
(A) Normalized intermediate scattering function $f_{xy}(k,t) = F_{xy}(k,t) / F_{xy}(k,0)$ (Eq.\ref{eq-isf}) of chain monomers at wave numbers $2\pi/k_{1} = 1.1$ $a$ and $2\pi/k_{3} = 3.6$ $a$. 
(B) Conformational relaxation time of chains $\tau_{i} = \int f_{xy}(k_{i},t) dt$ with different $\kappa$.
(C) Mean square displacement of trivalent cations, either trapped in the condensate or free in the bulk. Symbols have the same meanings as in (A).
(D) Diffusion coefficients of trapped trivalent cations (3+) and chain monomers.
}
\label{dynamics}
\end{figure}

\begin{figure*}[t]
\centering\includegraphics[width=0.6\linewidth]{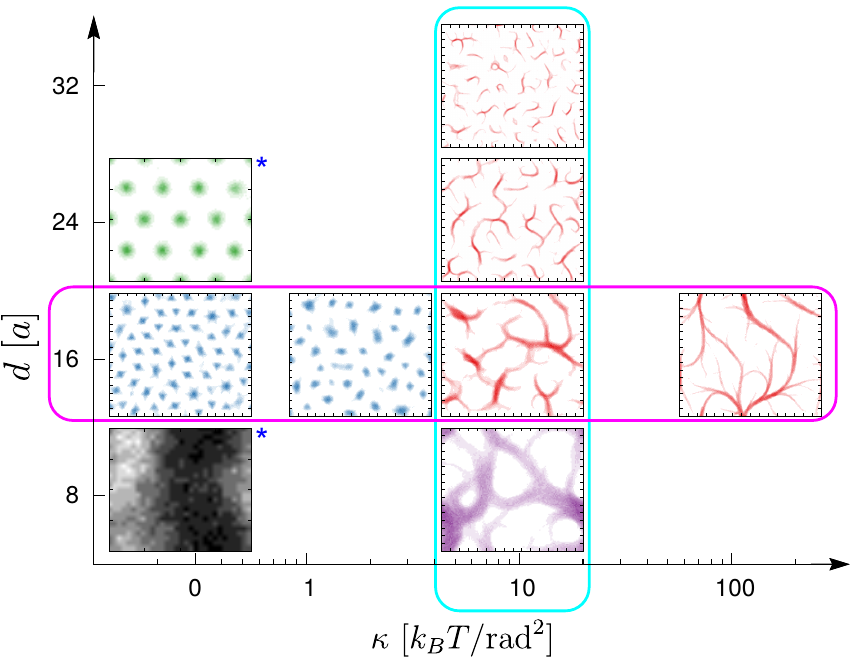}
\caption{
Time-averaged monomer density heat maps of PE brush condensates as a function of the bending rigidity parameter $\kappa$ and the grafting distance $d$.
Visually distinct morphologies are colored differently: homogeneous compact layer (black), octopus-like surface micelles (blue), 
single-chain tadpole-like condensate (green), dendritic domains and networks (red and purple).
Those labeled with asterisks (*) depict the simulation results of a smaller brush ($M=4\times4$) from our previous study \cite{2017Hyeon1579}.
}
\label{sum}
\end{figure*}

{\bf Dynamics of brush condensates. }
In order to quantify the dynamics of PE brush, 
we calculated the intermediate scattering function, which is the density-density time correlation function (van Hove correlation function) in Fourier domain,  
\begin{align}
F_{xy}(k,t) = \Big\langle \Big \langle \frac{1}{N_{m}} \sum_{m=1}^{N_{m}} e^{i \vec{k} \cdot \vec{r}_{m}(t+t_{0})} \sum_{n=1}^{N_{m}} e^{-i \vec{k} \cdot \vec{r}_{n}(t_{0})} \Big \rangle_{|\vec{k}|} \Big \rangle_{t_{0}} 
\label{eq-isf}
\end{align} 
where $\langle \langle \ldots \rangle_{|\vec{k}|} \rangle_{t_0}$ is an average over time $t_0$ and over the direction of a 2D wave vector $\vec{k}$ of magnitude $k$.
The dynamics of brush chain at different length scales can be probed in terms of  
$F_{xy}(k,t)$ evaluated at different $k$ ($k_{i} = 2\pi/r_{i}^{\ast}$ where $i=1$, 2, 3 and $r_{i}^{\ast}/a = 1.1$, 2.0, 3.6 are the positions of the three highest peaks in the radial distribution function of chain monomers (see Fig.~S4C,D)).
The normalized function $f_{xy}(k,t) = F_{xy}(k,t) / F_{xy}(k,0)$, with $k_{1}$ and $k_{3}$, are shown in Fig.~\ref{dynamics}A, 
and the corresponding mean relaxation time $\tau_{i} = \int_{0}^{\infty} f_{xy}(k_i,t) dt$ is presented in Fig.~\ref{dynamics}B. 
At a small length scale $k^{-1}_{1}$, $f_{xy}(k_1,t)$ decays to zero within the timescale of $t< 10 \tau_{\text{BD}}$, which implies that chain monomers are fluidic beyond this time scale. 
But, compared to octopus-like micelle with $\kappa=0$ $k_{B}T/\text{rad}^2$, 
the dendritic assembly made of brush chains with $\kappa=120$ $k_{B}T/\text{rad}^2$ displays $\sim 14$-fold slower relaxation profile of $f_{xy}(t)$.
The relaxation becomes much slower at larger length scale $k^{-1}_{3}$, 
and $\tau_{3}$ for rigid chain comprising the dendritic assembly is as long as our total simulation time ($\sim \mathcal{O}(10^3) \tau_{\text{BD}}$). 
We also notice that the ratio of relaxation times, $\eta_{i} = \tau_{i}(\kappa=120 \text{ }k_BT/\text{rad}^2) / \tau_{i}(\kappa=0 \text{ }k_BT/\text{rad}^2)$ at the three position of $r^{\ast}_i$ (with $i=1$, 2, 3) takes an order of $\eta_{3} > \eta_{2} > \eta_{1}$. 
This is expected because the contribution from inter-chain relaxation to the total relaxation time is higher at $r_{3}$ than at $r_{1}$. 
A tight and well aligned chain organization at $\kappa = 120$ $k_{B}T/\text{rad}^2$ further increases $\tau_{3}$ in comparison to $\tau_{1}$.
For the most rigid dendrite, $\tau_{3}$ is $\sim 60$-fold greater than that of the surface micelle formed by flexible PE brush.

Next, the mobility of trivalent cations, either trapped in the condensate (within $\lambda_{B}$ from the chains) or free in the bulk solution, were quantified using an ensemble- and time-averaged mean squared displacement, 
$\text{MSD}(t) = \langle \langle |\vec{r}_{i}(t+t_0) - \vec{r}_{i}(t_0)| \rangle_{t_0} \rangle$, as shown in Fig.~\ref{dynamics}C. 
When $\kappa \leq 5$ $k_{B}T/\text{rad}^2$, although trapped ions are mobile, 
MSD shows a long-time subdiffusive behavior because ions are confined in individual micelles \cite{2017Tirrell1497} (Supplementary Movie 1).
By contrast, for $\kappa > 10$ $k_{B}T/\text{rad}^2$, condensed ions can freely diffuse along the dendritic branches. 
As a result, MSD grows linearly with time.

The diffusion coefficient of trapped trivalent cation, estimated using $D = \text{MSD}(t) / 6 \Delta t$ for $\Delta t = 1\times 10^{3}$ to $1.5 \times 10^{3} \tau_{\text{BD}}$, is non-monotonic with $\kappa$. 
This change agrees with the change of the brush morphology where ions are confined.
In the micelle phase, micelle size grows with $\kappa$, which provides larger space for the trapped ions to navigate. 
In the dendrite phase, the effective attraction between neighboring chains, mediated by the counterions, increases with $\kappa$ (see Figs.~S2A, ~S4B) and  tightens the bundling of PE, which in turn reduces the mobility of the condensed ions. 
The trapped trivalent ions diffuse  $>$ 10-fold slower than those in the bulk, 
but still $\sim $ 100-fold faster than chain monomers in the dendritic assembly, 
even though the same value of bare diffusion coefficient was assumed for all ions and chain monomers.
The bundles of rigid chains form a network of ``highway", on which the condensed trivalent ions freely diffuse (Supplementary Movie 2).

\section{Discussion}

{\bf Effect of grafting density on the morphology of brush condensates. }  
The morphological transitions from the octopus-like surface micelles to the dendritic condensates are reminiscent of sol-to-gel transition.  
Analogous to gelation transition, the ``bond probability" $p$ can be tuned by changing either the chain stiffness ($\kappa$) or grafting distance ($d$). 
Below the gelation point ($p<p_c$) isolated domains are observed; and above the gelation point ($p>p_c$) the domains are all connected together, covering the entire space.  
We further performed simulations of a semiflexible brush, at $\kappa=10$ $k_{B}T/\text{rad}^2$, by varying the inter-chain spacing $d$ (see Fig.~\ref{sum} and Fig.~S6). 
Time-averaged monomer density heat maps of PE brushes $\langle \rho(x,y) \rangle$ (Fig.~\ref{sum}) visualize how the morphology of brush condensates changes as a function of the chain bending rigidity parameter $\kappa$ and the inter-chain spacing $d$. 

Notably, changes in $\kappa$ and $d$ display qualitatively different effects on the morphologies below and above the ``gelation point." 
At $d=16a$ with increasing $\kappa$ (panels enclosed by the magenta boundary in Fig.\ref{sum}), the initially sol-like micelles domain are percolated into gel-like dendritic pattern whose branches span the entire surface. 
In contrast, when the chain stiffness is fixed to $\kappa=10$ $k_BT/\text{rad}^2$ and grafting distance is varied from a large value ($d=32a$) to a small one ($d=8a$) (panels enclosed by the cyan boundary in Fig.\ref{sum}),  
the initial sol-like isolated domains are characterized by the heterogeneous condensates made of semiflexible chains, collapsed into toroids or rod-like bundles on site, not by the tadpole or octopus-like micelle condensates; and with decreasing $d$ the chains collapse and further assemble into a dendritic pattern and a non-uniform fractal-like meshwork layer. 
\\


{\bf Size of octopus-like surface micelle. } 
For octopus-like surface micelle, a scaling argument is developed based on equilibrium thermodynamics \cite{2017Hyeon1579}.  
The domain size is determined by the balance between the surface tension resulting from the counterion-mediated attraction 
and the elastic penalty to stretch the grafted chains to form a surface micelle. 
When $\kappa \leq 3$ $k_{B}T/\text{rad}^2$, $l_{p}$ is small enough to approximate the individual PE as a flexible chain with Kuhn length $2 l_{p}$. 
%
For an octopus-like domain containing $n$ chains within a surface area $\sim R_{c}^2 \simeq l_{p}^{2} (n N / l_{p})^{2 \nu}$, 
the surface energy is $F_{n,\text{surf}} = \xi k_{B}T l_{p}^{2} (n N a/ l_{p})^{2 \nu}$, 
where $\xi$ sets the scale of attraction between chain segments and $\nu$ is the Flory exponent. 
Meanwhile, the elastic penalty is $F_{n,\text{el}} = n k_{B} T R_{c}^2 / l^2_pN_{s} = n k_{B} T R_{c} /l_{p} = k_{B} T \sigma R_{c}^{3} / l_{p}$,
where $N_{s} = R_{c}/l_{p}$ is the number of statistical segments in each chain to be stretched to reach the micelle, 
and $\sigma = n/R_{c}^2$ is the chain grafting density.
The total free energy per area in the octopus-like condensate with $n$ arms is 
\begin{align}
\frac{f_{\text{octo}}}{k_{B}T} &= \frac{1}{k_BT}\frac{F_{n,\text{surf}}+F_{n,\text{el}}}{R_c^2}\nonumber\\
&= \frac{\xi (\sigma Na)^{2\nu} l_{p}^{2-2\nu}}{R_{c}^{2-4\nu}} + \frac{\sigma R_{c}}{l_p}. 
\label{eq-oct}
\end{align}
Minimization of $f_{octo}$ with respect to $R_{c}$ provides the micelle size corresponding to a minimum free energy 
$R_c^* \sim l_{p}^{\frac{3-2\nu}{3-4\nu}} (Na)^{\frac{2\nu}{3-4\nu}} \sigma^{\frac{2\nu-1}{3-4\nu}}$.
For $\nu = 1/3$, $R^*_{c}$ increases as $\sim l_{p}^{7/5}$ (thus with $\kappa$), until neighboring micelles is about to overlap.
Beyond this overlap point, the picture of isolated semispherical micelles no longer holds.
\\

{\bf Fractal dimension of dendritic condensate. } 
In the case of dendritic condensate, we found that $n(r)\sim r^{\mathcal{D}_f}$.   
In particular, $\mathcal{D}_{f} \approx 1.75$, observed at large $r$ ($r/a > 50$) was also reported in Bracha \emph{et al.}'s experiment \cite{2014BarZiv4945}. 
Incidentally, the morphology of aggregate changes depending on how trivalent salt is added \cite{2014BarZiv4945}.  
Thus, the formation of dendritic morphologies are effectively made under kinetic gelation rather than equilibrium one.  

%
%

The premise that the process of dendritic assembly is kinetically controlled guides the direction of our theoretical analysis.
Since the collapse is effectively irreversible and the bundles grow preferentially from the ``active front" of the preexisting domain \cite{1999Liu624},
we use the principle underlying the diffusion-limited aggregation \cite{1981Sander1400} (DLA) 
to explain the observed fractal dimension. 
DLA describes a far-from-equlibrium growth phenomenon, 
where each particle diffuse with a fractal dimension $d_{w}$ until it binds irreversibly to any particles of the aggregate. 
A generalized Honda-Toyoki-Matsushita mean-field approximation \cite{1984Kawasaki337,1986Kondo2618} suggests that, the fractal dimension of the aggregate is
\begin{equation}
\mathcal{D}_{\text{MF}} = \frac{d_s^2 + \eta (d_{w}-1)}{d_s + \eta (d_{w}-1)},
\label{eq-dla} 
\end{equation} 
where in the presence of long-range attractive interactions the probability of growth at a certain position is assumed to be proportional to the gradient of a scalar field (e.g. monomer density) as $\sim |\nabla\phi|^\eta$. 
For DLA ($\eta = 1$, $d_w=2$) in 2 dimension ($d_s=2$), Eq.\ref{eq-dla} gives $\mathcal{D}_{\text{MF}} =\mathcal{D}_{f,\text{DLA}} = 5/3$.
Numerical simulations report $\mathcal{D}_{f,\text{DLA}} = 1.71$ \cite{2016Dossetti,2017Dossetti3523}.  

DLA has also been exploited to explain the dynamics and aggregation of a 3D gel-like network formed by rigid PE chains in a poor solvent \cite{2017Asahi5991}. 
The fractal nature of the dendritic pattern may well be an outcome of premature quenching of brush configuration to condensates during the competition 
between the gain in energy upon aggregation and the entropic gain of chain fluctuations.
\\

\section{Concluding Remarks}
Collapse of the brush condensate into either surface micelles \cite{2017Tirrell1497} or a dendritic pattern \cite{2014BarZiv4945} is controlled by the chain flexibility.
Fundamental differences are found in the the dynamics of chains and condensed ions as well as in the microscopic chain arrangement.
The new insights into the link between the micro-scale details and brush morphology will be of great use to design material properties and understand biological functions of PE brushes. 

\section{Model and Methods}
\label{MaM}
{\bf Model and energy potential. } As in our previous study \cite{2017Hyeon1579}, we used a well tested coarse-grained model of strong polyelectrolyte (PE) brush \cite{2000Seidel2728,2003Stevens3855,2014Hsiao2900}. 
Total $M (= 16\times16)$ polymer chains were grafted to the uncharged surface of a 2D triangular lattice (Fig.~\ref{cfg}A). 
The lattice spacing $d$ was set to $16 a$, which is small enough to ensure the lateral overlap between neighboring chains, where $a$ is the diameter of chain monomers and ions.
Each chain consists of $N (= 80)$ negatively charged monomers and a neutral terminal monomer grafted to the surface. 
The simulation box has a dimension of 
$L_{x}\times L_{y}\times L_{z} = {(\sqrt{M} d)} \times {(\sqrt{3M} d/2)} \times {(2 N a)} = 256 a \times 128\sqrt{3} a \times 160 a$. 
Periodic boundary conditions were applied along the $x$ and $y$ axes, 
and impenetrable neutral walls were placed at $z=0$ and $2 N a$. 

We considered the following energy potentials to model a semiflexible PE brush in \emph{good} solvents with multivalent salts. 
First, the distance between the neighboring chain monomers was constrained by a finite extensible nonlinear elastic bond potential 
\begin{equation}
U_{bond}(r) = -\frac{k_{0} R_{0}^{2}}{2} \log\left(1-\frac{r^{2}}{R_{0}^{2}}\right),
\label{ub}
\end{equation}
with a spring constant $k_{0} = 5.83 ~k_{B}T/a^2$ and a maximum extensible bond length $R_{0} = 2a$.
Second, the chain stiffness was modulated with an angular potential 
\begin{equation}
U_{angle}(\theta) = \kappa (\theta - \pi)^{2},
\label{ua}
\end{equation}
where $\kappa$ is the bending rigidity parameter and $\theta$ is the angle between three consecutive monomers along the chain. 
Third, the excluded volume interaction was modeled in between ions and chain monomers 
by using the Weeks-Chandler-Andersen potential 
\begin{equation}
U_{excl}(r) = 4 \epsilon \left[\left(\frac{a}{r}\right)^{12}-\left(\frac{a}{r}\right)^{6}+\frac{1}{4}\right] \Theta(2^{1/6} a - r),
\label{ue}
\end{equation}
in which $\epsilon = 1 ~k_{B}T$ and $\Theta(\ldots)$ denotes a Heaviside step function.
Fourth, the Columbic interactions were assigned between charged particles $i$, $j$, which include both chain monomers and ions,   
\begin{equation}
U_{elec}(r) = \frac{k_{B} T \lambda_{B} z_{i} z_{j}}{r},
\label{uq}
\end{equation} 
where $z_{i,j}$ is the valence of charge. 
The Bjerrum length is defined as $\lambda_{B}=e^{2}/(4 \pi \epsilon_{0} \epsilon_{r} k_{B}T)$, 
where $\epsilon_{0}$ is the vacuum permittivity and $\epsilon_{r}$ is the relative dielectric constant of the solvent. 
Lastly, the confinement of the wall was considered to repel any monomer, 
that approaches the wall closer than $a/2$ such that
\begin{equation}
U_{wall}(z) = 4 \epsilon \left[\left(\frac{a}{z+\Delta}\right)^{12}-\left(\frac{a}{z+\Delta}\right)^{6}+\frac{1}{4}\right] \Theta(a/2 - z),
\label{uw}
\end{equation} 
with $\Delta = (2^{1/6}-1/2) a$. 
For simplicity, we assume the same diameter $a$ for all the ions and chain monomers.
For dsDNA, the mean bond length ${\langle b \rangle} = 1.1 a$ $(\approx a)$ in our model maps to 
the effective charge separation ($\approx 1.7$ {\AA}) along the chain.
Considering $\lambda_{B} = 7.1$ {\AA} in water at room temperature, we set $\lambda_{B}=4 a$ $(\approx 7.1/1.7\times a)$.
Since the focus of this study is on the effects of the bending rigidity of PE chain, $\kappa$ in Eq.\ref{ua} was adjusted in the range, $0\leq \kappa\leq 120$ $k_{B}T/\textrm{rad}^2$. 
\\

{\bf Simulation. } 
For conformational sampling of the brush, we integrated the Langevin equation at underdamped condition \cite{1992Thirumalai695}, 
\begin{align}
m\frac{d^2\vec{r}_i}{dt^2}=-\zeta_{\text{MD}}\frac{d\vec{r}_i}{dt}-\vec{\nabla}_{\vec{r}_i}U(\vec{r}_1,\vec{r}_2,\ldots)+\vec{\xi}(t), 
\end{align}
using a small friction coefficient $\zeta_{\text{MD}} = 0.1 m/\tau_{\text{MD}}$ and a time step $\delta t = 0.01 \tau_{\text{MD}}$, 
with the characteristic time scale $\tau_{\text{MD}} = (ma^2/{\epsilon})^{1/2}$. 
We started from an initial configuration where polymer chains were vertically stretched, 
and monovalent counterions were homogeneously distributed in the brush region.
This salt-free brush was first equilibrated for the time of $10^4$ $\tau_{\text{MD}}$, then trivalent cations at a 1:3 stoichiometric concentration ratio 
with respect to the polyelectrolyte charges \cite{2017Hyeon1579} were randomly added together with its monovalent coions (anions) into the brush-free zone (see Fig.~\ref{cfg}A). 
Depending on the value of $\kappa$, 
trivalent cations induce an immediate collapse or bundling of neighboring chains in the brush. 
In the latter case, an intermediate bundle either merges into a thicker one with other bundles nearby, 
or collapses onto the grafting surface \emph{irreversibly}.
For stiff chains with $\kappa = 60$ or $120$ $k_{B}T/\textrm{rad}^2$, it takes longer than $ 10^{6}$ $\tau_{\text{MD}}$ 
before the whole brush collapses and the mean height of chains reaches the steady state. 
Production runs was generated further for $5\times10^{4} \tau_{\text{MD}}$.  
Brush configurations were collected every $50 \tau_{\text{MD}}$ for the analysis of static properties.
Unless stated otherwise, all the conformational properties reported here were averaged over the ensemble of trajectories.

To probe the dynamics of condensates, 
we performed Brownian dynamics (BD) simulations by integrating the following equation of motion 
\begin{equation}
\frac{d \vec{r}_{i}}{dt} = - \frac{D_{i0}}{k_B T}{\vec{\nabla}}_{\vec{r}_{i}} U(\vec{r}_1,...,\vec{r}_N) +\vec{R}_{i}(t),
\end{equation}
where $D_{i0}$ is the bare diffusion coefficient of the $i$-th particle, 
and $\vec{R}_{i}(t)$ is the Gaussian noise satisfying $\langle\vec{R}_i(t)\rangle=0$ and  
$\langle \vec{R}_{i}(t) \cdot \vec{R}_{j}(t')\rangle = 6 D_{i0} \delta_{ij} \delta(t-t')$. 
$D_{i0}$ was estimated via $k_{B}T / 6 \pi \eta R$, where $\eta = 0.89\times 10^{3}$ Pa$\cdot$s is the viscosity of water 
and $R$ is the hydration radius of all the particles. 
We chose an integration time step $\delta t_{\text{BD}} = 2 \times 10^{-4} \tau_{\text{BD}}$ 
with the Brownian time $\tau_{\text{BD}} = a^{2}/D_{i0}$ ($\sim 4$ ns, assuming that $R \sim 10$ {\AA}).
Starting from the last configuration of brush in MD simulations, the BD simulation was performed for $4 \times 10^{3} \tau_{\text{BD}}$.
Simulations were all carried out by using ESPResSo 3.3.1 package \cite{2006Holm704,2013Holm1}. More details can be found in Ref.\cite{2017Hyeon1579}. 
\\

{\bf Apparent persistence length of brush chain. } 
By using a simplifying assumption that as an isolated semiflexible chain the correlation between bond vectors exponentially decays with the their separation along the chain ($g(s)=\langle \vec{u}_i\cdot\vec{u}_{i+s}\rangle \sim e^{-s/l_p}$, where $\vec{u}_i=(\vec{r}_{i+1} - \vec{r}_{i}) / |\vec{r}_{i+1} - \vec{r}_{i}|$) (Fig.~S7),  
we quantified an ``apparent" persistent length $l_{p}$. 




%
%

%

\section{Supplementary Material}
Supplementary material contains the Supplementary Figures S1 -- S7 and Supplementary Movies 1 and 2. 

\begin{acknowledgments}
We thank the Center for Advanced Computation in KIAS for providing computing resources.  
\end{acknowledgments}


\begin{thebibliography}{63}%
\makeatletter
\providecommand \@ifxundefined [1]{%
 \@ifx{#1\undefined}
}%
\providecommand \@ifnum [1]{%
 \ifnum #1\expandafter \@firstoftwo
 \else \expandafter \@secondoftwo
 \fi
}%
\providecommand \@ifx [1]{%
 \ifx #1\expandafter \@firstoftwo
 \else \expandafter \@secondoftwo
 \fi
}%
\providecommand \natexlab [1]{#1}%
\providecommand \enquote  [1]{``#1''}%
\providecommand \bibnamefont  [1]{#1}%
\providecommand \bibfnamefont [1]{#1}%
\providecommand \citenamefont [1]{#1}%
\providecommand \href@noop [0]{\@secondoftwo}%
\providecommand \href [0]{\begingroup \@sanitize@url \@href}%
\providecommand \@href[1]{\@@startlink{#1}\@@href}%
\providecommand \@@href[1]{\endgroup#1\@@endlink}%
\providecommand \@sanitize@url [0]{\catcode `\\12\catcode `\$12\catcode
  `\&12\catcode `\#12\catcode `\^12\catcode `\_12\catcode `\%12\relax}%
\providecommand \@@startlink[1]{}%
\providecommand \@@endlink[0]{}%
\providecommand \url  [0]{\begingroup\@sanitize@url \@url }%
\providecommand \@url [1]{\endgroup\@href {#1}{\urlprefix }}%
\providecommand \urlprefix  [0]{URL }%
\providecommand \Eprint [0]{\href }%
\providecommand \doibase [0]{http://dx.doi.org/}%
\providecommand \selectlanguage [0]{\@gobble}%
\providecommand \bibinfo  [0]{\@secondoftwo}%
\providecommand \bibfield  [0]{\@secondoftwo}%
\providecommand \translation [1]{[#1]}%
\providecommand \BibitemOpen [0]{}%
\providecommand \bibitemStop [0]{}%
\providecommand \bibitemNoStop [0]{.\EOS\space}%
\providecommand \EOS [0]{\spacefactor3000\relax}%
\providecommand \BibitemShut  [1]{\csname bibitem#1\endcsname}%
\let\auto@bib@innerbib\@empty
\bibitem [{\citenamefont {Oosawa}(1971)}]{OosawaBook}%
  \BibitemOpen
  \bibfield  {author} {\bibinfo {author} {\bibfnamefont {F.}~\bibnamefont
  {Oosawa}},\ }\href@noop {} {\emph {\bibinfo {title} {Polyelectrolytes}}}\
  (\bibinfo  {publisher} {Marcel Dekker, Inc.},\ \bibinfo {year}
  {1971})\BibitemShut {NoStop}%
\bibitem [{\citenamefont {Skolnick}\ and\ \citenamefont
  {Fixman}(1977)}]{SkolnickMacro77}%
  \BibitemOpen
  \bibfield  {author} {\bibinfo {author} {\bibfnamefont {J.}~\bibnamefont
  {Skolnick}}\ and\ \bibinfo {author} {\bibfnamefont {M.}~\bibnamefont
  {Fixman}},\ }\bibfield  {title} {\enquote {\bibinfo {title} {Electrostatic
  persistence length of a wormlike polyelectrolyte},}\ }\href@noop {}
  {\bibfield  {journal} {\bibinfo  {journal} {Macromolecules}\ }\textbf
  {\bibinfo {volume} {10}},\ \bibinfo {pages} {944--948} (\bibinfo {year}
  {1977})}\BibitemShut {NoStop}%
\bibitem [{\citenamefont {Barrat}\ and\ \citenamefont
  {Joanny}(1993)}]{Barrat93EL}%
  \BibitemOpen
  \bibfield  {author} {\bibinfo {author} {\bibfnamefont {J.~L.}\ \bibnamefont
  {Barrat}}\ and\ \bibinfo {author} {\bibfnamefont {J.~F.}\ \bibnamefont
  {Joanny}},\ }\bibfield  {title} {\enquote {\bibinfo {title} {Persistence
  length of polyelectrolyte chains},}\ }\href@noop {} {\bibfield  {journal}
  {\bibinfo  {journal} {Europhys. Lett.}\ }\textbf {\bibinfo {volume} {24}},\
  \bibinfo {pages} {333} (\bibinfo {year} {1993})}\BibitemShut {NoStop}%
\bibitem [{\citenamefont {Ha}\ and\ \citenamefont
  {Thirumalai}(1995)}]{ha1995macromolecules}%
  \BibitemOpen
  \bibfield  {author} {\bibinfo {author} {\bibfnamefont {B.-Y.}\ \bibnamefont
  {Ha}}\ and\ \bibinfo {author} {\bibfnamefont {D.}~\bibnamefont
  {Thirumalai}},\ }\bibfield  {title} {\enquote {\bibinfo {title}
  {Electrostatic persistence length of a polyelectrolyte chain},}\ }\href@noop
  {} {\bibfield  {journal} {\bibinfo  {journal} {Macromolecules}\ }\textbf
  {\bibinfo {volume} {28}},\ \bibinfo {pages} {577--581} (\bibinfo {year}
  {1995})}\BibitemShut {NoStop}%
\bibitem [{\citenamefont {Schiessel}(1999)}]{schiessel1999macromolecules}%
  \BibitemOpen
  \bibfield  {author} {\bibinfo {author} {\bibfnamefont {H.}~\bibnamefont
  {Schiessel}},\ }\bibfield  {title} {\enquote {\bibinfo {title} {Counterion
  condensation on flexible polyelectrolytes: dependence on ionic strength and
  chain concentration},}\ }\href@noop {} {\bibfield  {journal} {\bibinfo
  {journal} {Macromolecules}\ }\textbf {\bibinfo {volume} {32}},\ \bibinfo
  {pages} {5673--5680} (\bibinfo {year} {1999})}\BibitemShut {NoStop}%
\bibitem [{\citenamefont {Caliskan}\ \emph {et~al.}(2005)\citenamefont
  {Caliskan}, \citenamefont {Hyeon}, \citenamefont {Perez-Salas}, \citenamefont
  {Briber}, \citenamefont {Woodson},\ and\ \citenamefont
  {Thirumalai}}]{Caliskan05PRL}%
  \BibitemOpen
  \bibfield  {author} {\bibinfo {author} {\bibfnamefont {G.}~\bibnamefont
  {Caliskan}}, \bibinfo {author} {\bibfnamefont {C.}~\bibnamefont {Hyeon}},
  \bibinfo {author} {\bibfnamefont {U.}~\bibnamefont {Perez-Salas}}, \bibinfo
  {author} {\bibfnamefont {R.~M.}\ \bibnamefont {Briber}}, \bibinfo {author}
  {\bibfnamefont {S.~A.}\ \bibnamefont {Woodson}}, \ and\ \bibinfo {author}
  {\bibfnamefont {D.}~\bibnamefont {Thirumalai}},\ }\bibfield  {title}
  {\enquote {\bibinfo {title} {{Persistence Length Changes Dramatically as RNA
  Folds}},}\ }\href@noop {} {\bibfield  {journal} {\bibinfo  {journal} {Phys.
  Rev. Lett.}\ }\textbf {\bibinfo {volume} {95}},\ \bibinfo {pages} {268303}
  (\bibinfo {year} {2005})}\BibitemShut {NoStop}%
\bibitem [{\citenamefont {Moghaddam}\ \emph {et~al.}(2009)\citenamefont
  {Moghaddam}, \citenamefont {Caliskan}, \citenamefont {Chauhan}, \citenamefont
  {Hyeon}, \citenamefont {Briber}, \citenamefont {Thirumalai},\ and\
  \citenamefont {Woodson}}]{moghaddam09JMB}%
  \BibitemOpen
  \bibfield  {author} {\bibinfo {author} {\bibfnamefont {S.}~\bibnamefont
  {Moghaddam}}, \bibinfo {author} {\bibfnamefont {G.}~\bibnamefont {Caliskan}},
  \bibinfo {author} {\bibfnamefont {S.}~\bibnamefont {Chauhan}}, \bibinfo
  {author} {\bibfnamefont {C.}~\bibnamefont {Hyeon}}, \bibinfo {author}
  {\bibfnamefont {R.}~\bibnamefont {Briber}}, \bibinfo {author} {\bibfnamefont
  {D.}~\bibnamefont {Thirumalai}}, \ and\ \bibinfo {author} {\bibfnamefont
  {S.}~\bibnamefont {Woodson}},\ }\bibfield  {title} {\enquote {\bibinfo
  {title} {{Metal ion dependence of cooperative collapse transitions in
  RNA}},}\ }\href@noop {} {\bibfield  {journal} {\bibinfo  {journal} {J. Mol.
  Biol.}\ }\textbf {\bibinfo {volume} {393}},\ \bibinfo {pages} {753--764}
  (\bibinfo {year} {2009})}\BibitemShut {NoStop}%
\bibitem [{\citenamefont {Liu}\ and\ \citenamefont {Hyeon}(2016)}]{liu2016BJ}%
  \BibitemOpen
  \bibfield  {author} {\bibinfo {author} {\bibfnamefont {L.}~\bibnamefont
  {Liu}}\ and\ \bibinfo {author} {\bibfnamefont {C.}~\bibnamefont {Hyeon}},\
  }\bibfield  {title} {\enquote {\bibinfo {title} {Contact statistics highlight
  distinct organizing principles of proteins and rna},}\ }\href@noop {}
  {\bibfield  {journal} {\bibinfo  {journal} {Biophys. J.}\ }\textbf {\bibinfo
  {volume} {110}},\ \bibinfo {pages} {2320--2327} (\bibinfo {year}
  {2016})}\BibitemShut {NoStop}%
\bibitem [{\citenamefont {Emanuel}\ \emph {et~al.}(2009)\citenamefont
  {Emanuel}, \citenamefont {Radja}, \citenamefont {Henriksson},\ and\
  \citenamefont {Schiessel}}]{emanuel2009PhysBiol}%
  \BibitemOpen
  \bibfield  {author} {\bibinfo {author} {\bibfnamefont {M.}~\bibnamefont
  {Emanuel}}, \bibinfo {author} {\bibfnamefont {N.~H.}\ \bibnamefont {Radja}},
  \bibinfo {author} {\bibfnamefont {A.}~\bibnamefont {Henriksson}}, \ and\
  \bibinfo {author} {\bibfnamefont {H.}~\bibnamefont {Schiessel}},\ }\bibfield
  {title} {\enquote {\bibinfo {title} {{The physics behind the larger scale
  organization of DNA in eukaryotes}},}\ }\href@noop {} {\bibfield  {journal}
  {\bibinfo  {journal} {Phys. Biol.}\ }\textbf {\bibinfo {volume} {6}},\
  \bibinfo {pages} {025008} (\bibinfo {year} {2009})}\BibitemShut {NoStop}%
\bibitem [{\citenamefont {Needleman}\ \emph {et~al.}(2004)\citenamefont
  {Needleman}, \citenamefont {Ojeda-Lopez}, \citenamefont {Raviv},
  \citenamefont {Miller}, \citenamefont {Wilson},\ and\ \citenamefont
  {Safinya}}]{needleman2004PNAS}%
  \BibitemOpen
  \bibfield  {author} {\bibinfo {author} {\bibfnamefont {D.~J.}\ \bibnamefont
  {Needleman}}, \bibinfo {author} {\bibfnamefont {M.~A.}\ \bibnamefont
  {Ojeda-Lopez}}, \bibinfo {author} {\bibfnamefont {U.}~\bibnamefont {Raviv}},
  \bibinfo {author} {\bibfnamefont {H.~P.}\ \bibnamefont {Miller}}, \bibinfo
  {author} {\bibfnamefont {L.}~\bibnamefont {Wilson}}, \ and\ \bibinfo {author}
  {\bibfnamefont {C.~R.}\ \bibnamefont {Safinya}},\ }\bibfield  {title}
  {\enquote {\bibinfo {title} {Higher-order assembly of microtubules by
  counterions: from hexagonal bundles to living necklaces},}\ }\href@noop {}
  {\bibfield  {journal} {\bibinfo  {journal} {Proc. Natl. Acad. Sci. U. S. A.}\
  }\textbf {\bibinfo {volume} {101}},\ \bibinfo {pages} {16099--16103}
  (\bibinfo {year} {2004})}\BibitemShut {NoStop}%
\bibitem [{\citenamefont {Hud}\ and\ \citenamefont
  {Vilfan}(2005)}]{hud2005ARBBS}%
  \BibitemOpen
  \bibfield  {author} {\bibinfo {author} {\bibfnamefont {N.~V.}\ \bibnamefont
  {Hud}}\ and\ \bibinfo {author} {\bibfnamefont {I.~D.}\ \bibnamefont
  {Vilfan}},\ }\bibfield  {title} {\enquote {\bibinfo {title} {Toroidal dna
  condensates: unraveling the fine structure and the role of nucleation in
  determining size},}\ }\href@noop {} {\bibfield  {journal} {\bibinfo
  {journal} {Annu. Rev. Biophys. Biomol. Struct.}\ }\textbf {\bibinfo {volume}
  {34}},\ \bibinfo {pages} {295--318} (\bibinfo {year} {2005})}\BibitemShut
  {NoStop}%
\bibitem [{\citenamefont {Kang}\ \emph {et~al.}(2015)\citenamefont {Kang},
  \citenamefont {Yoon}, \citenamefont {Thirumalai},\ and\ \citenamefont
  {Hyeon}}]{2015Hyeon198102}%
  \BibitemOpen
  \bibfield  {author} {\bibinfo {author} {\bibfnamefont {H.}~\bibnamefont
  {Kang}}, \bibinfo {author} {\bibfnamefont {Y.-G.}\ \bibnamefont {Yoon}},
  \bibinfo {author} {\bibfnamefont {D.}~\bibnamefont {Thirumalai}}, \ and\
  \bibinfo {author} {\bibfnamefont {C.}~\bibnamefont {Hyeon}},\ }\bibfield
  {title} {\enquote {\bibinfo {title} {Confinement-induced glassy dynamics in a
  model for chromosome organization},}\ }\href {\doibase
  10.1103/PhysRevLett.115.198102} {\bibfield  {journal} {\bibinfo  {journal}
  {Phys. Rev. Lett.}\ }\textbf {\bibinfo {volume} {115}},\ \bibinfo {pages}
  {198102} (\bibinfo {year} {2015})}\BibitemShut {NoStop}%
\bibitem [{\citenamefont {Martin}\ \emph {et~al.}(2000)\citenamefont {Martin},
  \citenamefont {Davies}, \citenamefont {Rackstraw}, \citenamefont {Roberts},
  \citenamefont {Stolnik}, \citenamefont {Tendler},\ and\ \citenamefont
  {Williams}}]{2000Williams106}%
  \BibitemOpen
  \bibfield  {author} {\bibinfo {author} {\bibfnamefont {A.}~\bibnamefont
  {Martin}}, \bibinfo {author} {\bibfnamefont {M.}~\bibnamefont {Davies}},
  \bibinfo {author} {\bibfnamefont {B.}~\bibnamefont {Rackstraw}}, \bibinfo
  {author} {\bibfnamefont {C.}~\bibnamefont {Roberts}}, \bibinfo {author}
  {\bibfnamefont {S.}~\bibnamefont {Stolnik}}, \bibinfo {author} {\bibfnamefont
  {S.}~\bibnamefont {Tendler}}, \ and\ \bibinfo {author} {\bibfnamefont
  {P.}~\bibnamefont {Williams}},\ }\bibfield  {title} {\enquote {\bibinfo
  {title} {Observation of {DNA}-polymer condensate formation in real time at a
  molecular level},}\ }\href {\doibase
  https://doi.org/10.1016/S0014-5793(00)01894-9} {\bibfield  {journal}
  {\bibinfo  {journal} {FEBS Letters}\ }\textbf {\bibinfo {volume} {480}},\
  \bibinfo {pages} {106 -- 112} (\bibinfo {year} {2000})}\BibitemShut {NoStop}%
\bibitem [{\citenamefont {Vilfan}\ \emph {et~al.}(2006)\citenamefont {Vilfan},
  \citenamefont {Conwell}, \citenamefont {Sarkar},\ and\ \citenamefont
  {Hud}}]{2006Hud8174}%
  \BibitemOpen
  \bibfield  {author} {\bibinfo {author} {\bibfnamefont {I.~D.}\ \bibnamefont
  {Vilfan}}, \bibinfo {author} {\bibfnamefont {C.~C.}\ \bibnamefont {Conwell}},
  \bibinfo {author} {\bibfnamefont {T.}~\bibnamefont {Sarkar}}, \ and\ \bibinfo
  {author} {\bibfnamefont {N.~V.}\ \bibnamefont {Hud}},\ }\bibfield  {title}
  {\enquote {\bibinfo {title} {Time study of {DNA} condensate morphology:
  Implications regarding the nucleation, growth, and equilibrium populations of
  toroids and rods},}\ }\href {\doibase 10.1021/bi060396c} {\bibfield
  {journal} {\bibinfo  {journal} {Biochemistry}\ }\textbf {\bibinfo {volume}
  {45}},\ \bibinfo {pages} {8174--8183} (\bibinfo {year} {2006})}\BibitemShut
  {NoStop}%
\bibitem [{\citenamefont {Stevens}(2001)}]{2001Stevens130}%
  \BibitemOpen
  \bibfield  {author} {\bibinfo {author} {\bibfnamefont {M.~J.}\ \bibnamefont
  {Stevens}},\ }\bibfield  {title} {\enquote {\bibinfo {title} {Simple
  simulations of {DNA} condensation},}\ }\href {\doibase
  http://dx.doi.org/10.1016/S0006-3495(01)76000-6} {\bibfield  {journal}
  {\bibinfo  {journal} {Biophys. J.}\ }\textbf {\bibinfo {volume} {80}},\
  \bibinfo {pages} {130 -- 139} (\bibinfo {year} {2001})}\BibitemShut {NoStop}%
\bibitem [{\citenamefont {Lee}\ and\ \citenamefont
  {Thirumalai}(2001)}]{2001Lee3446}%
  \BibitemOpen
  \bibfield  {author} {\bibinfo {author} {\bibfnamefont {N.}~\bibnamefont
  {Lee}}\ and\ \bibinfo {author} {\bibfnamefont {D.}~\bibnamefont
  {Thirumalai}},\ }\bibfield  {title} {\enquote {\bibinfo {title} {Dynamics of
  collapse of flexible polyelectrolytes in poor solvents},}\ }\href {\doibase
  10.1021/ma001604q} {\bibfield  {journal} {\bibinfo  {journal}
  {Macromolecules}\ }\textbf {\bibinfo {volume} {34}},\ \bibinfo {pages}
  {3446--3457} (\bibinfo {year} {2001})}\BibitemShut {NoStop}%
\bibitem [{\citenamefont {Ou}\ and\ \citenamefont
  {Muthukumar}(2005)}]{2005Muthukumar074905}%
  \BibitemOpen
  \bibfield  {author} {\bibinfo {author} {\bibfnamefont {Z.}~\bibnamefont
  {Ou}}\ and\ \bibinfo {author} {\bibfnamefont {M.}~\bibnamefont
  {Muthukumar}},\ }\bibfield  {title} {\enquote {\bibinfo {title} {Langevin
  dynamics of semiflexible polyelectrolytes: Rod-toroid-globule-coil structures
  and counterion distribution},}\ }\href {\doibase 10.1063/1.1940054}
  {\bibfield  {journal} {\bibinfo  {journal} {J. Chem. Phys.}\ }\textbf
  {\bibinfo {volume} {123}},\ \bibinfo {pages} {074905} (\bibinfo {year}
  {2005})}\BibitemShut {NoStop}%
\bibitem [{\citenamefont {Seaton}\ \emph {et~al.}(2013)\citenamefont {Seaton},
  \citenamefont {Schnabel}, \citenamefont {Landau},\ and\ \citenamefont
  {Bachmann}}]{2013Bachmann028103}%
  \BibitemOpen
  \bibfield  {author} {\bibinfo {author} {\bibfnamefont {D.~T.}\ \bibnamefont
  {Seaton}}, \bibinfo {author} {\bibfnamefont {S.}~\bibnamefont {Schnabel}},
  \bibinfo {author} {\bibfnamefont {D.~P.}\ \bibnamefont {Landau}}, \ and\
  \bibinfo {author} {\bibfnamefont {M.}~\bibnamefont {Bachmann}},\ }\bibfield
  {title} {\enquote {\bibinfo {title} {From flexible to stiff: Systematic
  analysis of structural phases for single semiflexible polymers},}\ }\href
  {\doibase 10.1103/PhysRevLett.110.028103} {\bibfield  {journal} {\bibinfo
  {journal} {Phys. Rev. Lett.}\ }\textbf {\bibinfo {volume} {110}},\ \bibinfo
  {pages} {028103} (\bibinfo {year} {2013})}\BibitemShut {NoStop}%
\bibitem [{\citenamefont {Hoang}\ \emph {et~al.}(2014)\citenamefont {Hoang},
  \citenamefont {Giacometti}, \citenamefont {Podgornik}, \citenamefont
  {Nguyen}, \citenamefont {Banavar},\ and\ \citenamefont
  {Maritan}}]{2014Maritan064902}%
  \BibitemOpen
  \bibfield  {author} {\bibinfo {author} {\bibfnamefont {T.~X.}\ \bibnamefont
  {Hoang}}, \bibinfo {author} {\bibfnamefont {A.}~\bibnamefont {Giacometti}},
  \bibinfo {author} {\bibfnamefont {R.}~\bibnamefont {Podgornik}}, \bibinfo
  {author} {\bibfnamefont {N.~T.~T.}\ \bibnamefont {Nguyen}}, \bibinfo {author}
  {\bibfnamefont {J.~R.}\ \bibnamefont {Banavar}}, \ and\ \bibinfo {author}
  {\bibfnamefont {A.}~\bibnamefont {Maritan}},\ }\bibfield  {title} {\enquote
  {\bibinfo {title} {From toroidal to rod-like condensates of semiflexible
  polymers},}\ }\href {\doibase 10.1063/1.4863996} {\bibfield  {journal}
  {\bibinfo  {journal} {J. Chem. Phys.}\ }\textbf {\bibinfo {volume} {140}},\
  \bibinfo {pages} {064902} (\bibinfo {year} {2014})}\BibitemShut {NoStop}%
\bibitem [{\citenamefont {Morrison}\ and\ \citenamefont
  {Thirumalai}(2009)}]{morrison2009PRE}%
  \BibitemOpen
  \bibfield  {author} {\bibinfo {author} {\bibfnamefont {G.}~\bibnamefont
  {Morrison}}\ and\ \bibinfo {author} {\bibfnamefont {D.}~\bibnamefont
  {Thirumalai}},\ }\bibfield  {title} {\enquote {\bibinfo {title} {Semiflexible
  chains in confined spaces},}\ }\href@noop {} {\bibfield  {journal} {\bibinfo
  {journal} {Phys. Rev. E.}\ }\textbf {\bibinfo {volume} {79}},\ \bibinfo
  {pages} {011924} (\bibinfo {year} {2009})}\BibitemShut {NoStop}%
\bibitem [{\citenamefont {Spakowitz}\ and\ \citenamefont
  {Wang}(2003)}]{spakowitz2003PRL}%
  \BibitemOpen
  \bibfield  {author} {\bibinfo {author} {\bibfnamefont {A.~J.}\ \bibnamefont
  {Spakowitz}}\ and\ \bibinfo {author} {\bibfnamefont {Z.-G.}\ \bibnamefont
  {Wang}},\ }\bibfield  {title} {\enquote {\bibinfo {title} {Semiflexible
  polymer confined to a spherical surface},}\ }\href@noop {} {\bibfield
  {journal} {\bibinfo  {journal} {Phys, Rev. Lett.}\ }\textbf {\bibinfo
  {volume} {91}},\ \bibinfo {pages} {166102} (\bibinfo {year}
  {2003})}\BibitemShut {NoStop}%
\bibitem [{\citenamefont {Hud}\ and\ \citenamefont
  {Downing}(2001)}]{2001Kenneth14925}%
  \BibitemOpen
  \bibfield  {author} {\bibinfo {author} {\bibfnamefont {N.~V.}\ \bibnamefont
  {Hud}}\ and\ \bibinfo {author} {\bibfnamefont {K.~H.}\ \bibnamefont
  {Downing}},\ }\bibfield  {title} {\enquote {\bibinfo {title} {Cryoelectron
  microscopy of $\lambda$ phage {DNA} condensates in vitreous ice: The fine
  structure of {DNA} toroids},}\ }\href {\doibase 10.1073/pnas.261560398}
  {\bibfield  {journal} {\bibinfo  {journal} {Proc. Natl. Acad. Sci. USA}\
  }\textbf {\bibinfo {volume} {98}},\ \bibinfo {pages} {14925--14930} (\bibinfo
  {year} {2001})}\BibitemShut {NoStop}%
\bibitem [{\citenamefont {Leforestier}(2013)}]{2013Leforestier201}%
  \BibitemOpen
  \bibfield  {author} {\bibinfo {author} {\bibfnamefont {A.}~\bibnamefont
  {Leforestier}},\ }\bibfield  {title} {\enquote {\bibinfo {title}
  {Polymorphism of {DNA} conformation inside the bacteriophage capsid},}\
  }\href {\doibase 10.1007/s10867-013-9315-y} {\bibfield  {journal} {\bibinfo
  {journal} {J. Biol. Phys.}\ }\textbf {\bibinfo {volume} {39}},\ \bibinfo
  {pages} {201--213} (\bibinfo {year} {2013})}\BibitemShut {NoStop}%
\bibitem [{\citenamefont {Berndsen}\ \emph {et~al.}(2014)\citenamefont
  {Berndsen}, \citenamefont {Keller}, \citenamefont {Grimes}, \citenamefont
  {Jardine},\ and\ \citenamefont {Smith}}]{Berndsen14PNAS}%
  \BibitemOpen
  \bibfield  {author} {\bibinfo {author} {\bibfnamefont {Z.~T.}\ \bibnamefont
  {Berndsen}}, \bibinfo {author} {\bibfnamefont {N.}~\bibnamefont {Keller}},
  \bibinfo {author} {\bibfnamefont {S.}~\bibnamefont {Grimes}}, \bibinfo
  {author} {\bibfnamefont {P.~J.}\ \bibnamefont {Jardine}}, \ and\ \bibinfo
  {author} {\bibfnamefont {D.~E.}\ \bibnamefont {Smith}},\ }\bibfield  {title}
  {\enquote {\bibinfo {title} {{Nonequilibrium dynamics and ultraslow
  relaxation of confined DNA during viral packaging}},}\ }\href@noop {}
  {\bibfield  {journal} {\bibinfo  {journal} {Proc. Natl. Acad. Sci. U. S. A.}\
  }\textbf {\bibinfo {volume} {111}},\ \bibinfo {pages} {8345--8350} (\bibinfo
  {year} {2014})}\BibitemShut {NoStop}%
\bibitem [{\citenamefont {Allahverdi}\ \emph {et~al.}(2015)\citenamefont
  {Allahverdi}, \citenamefont {Chen}, \citenamefont {Korolev},\ and\
  \citenamefont {Nordenski{\"o}ld}}]{2015Nordenskiold8512}%
  \BibitemOpen
  \bibfield  {author} {\bibinfo {author} {\bibfnamefont {A.}~\bibnamefont
  {Allahverdi}}, \bibinfo {author} {\bibfnamefont {Q.}~\bibnamefont {Chen}},
  \bibinfo {author} {\bibfnamefont {N.}~\bibnamefont {Korolev}}, \ and\
  \bibinfo {author} {\bibfnamefont {L.}~\bibnamefont {Nordenski{\"o}ld}},\
  }\bibfield  {title} {\enquote {\bibinfo {title} {Chromatin compaction under
  mixed salt conditions: Opposite effects of sodium and potassium ions on
  nucleosome array folding},}\ }\href {\doibase 10.1038/srep08512} {\bibfield
  {journal} {\bibinfo  {journal} {Sci. Rep.}\ }\textbf {\bibinfo {volume}
  {5}},\ \bibinfo {pages} {8512} (\bibinfo {year} {2015})}\BibitemShut
  {NoStop}%
\bibitem [{\citenamefont {Yoo}\ \emph {et~al.}(2016)\citenamefont {Yoo},
  \citenamefont {Kim}, \citenamefont {Aksimentiev},\ and\ \citenamefont
  {Ha}}]{yoo2016NatComm}%
  \BibitemOpen
  \bibfield  {author} {\bibinfo {author} {\bibfnamefont {J.}~\bibnamefont
  {Yoo}}, \bibinfo {author} {\bibfnamefont {H.}~\bibnamefont {Kim}}, \bibinfo
  {author} {\bibfnamefont {A.}~\bibnamefont {Aksimentiev}}, \ and\ \bibinfo
  {author} {\bibfnamefont {T.}~\bibnamefont {Ha}},\ }\bibfield  {title}
  {\enquote {\bibinfo {title} {Direct evidence for sequence-dependent
  attraction between double-stranded dna controlled by methylation},}\
  }\href@noop {} {\bibfield  {journal} {\bibinfo  {journal} {Nat. Commun.}\
  }\textbf {\bibinfo {volume} {7}},\ \bibinfo {pages} {11045} (\bibinfo {year}
  {2016})}\BibitemShut {NoStop}%
\bibitem [{\citenamefont {Pincus}(1991)}]{1991Pincus2912}%
  \BibitemOpen
  \bibfield  {author} {\bibinfo {author} {\bibfnamefont {P.}~\bibnamefont
  {Pincus}},\ }\bibfield  {title} {\enquote {\bibinfo {title} {Colloid
  stabilization with grafted polyelectrolytes},}\ }\href {\doibase
  10.1021/ma00010a043} {\bibfield  {journal} {\bibinfo  {journal}
  {Macromolecules}\ }\textbf {\bibinfo {volume} {24}},\ \bibinfo {pages}
  {2912--2919} (\bibinfo {year} {1991})}\BibitemShut {NoStop}%
\bibitem [{\citenamefont {Israels}\ \emph {et~al.}(1994)\citenamefont
  {Israels}, \citenamefont {Leermakers}, \citenamefont {Fleer},\ and\
  \citenamefont {Zhulina}}]{1994Zhulina3249}%
  \BibitemOpen
  \bibfield  {author} {\bibinfo {author} {\bibfnamefont {R.}~\bibnamefont
  {Israels}}, \bibinfo {author} {\bibfnamefont {F.~A.~M.}\ \bibnamefont
  {Leermakers}}, \bibinfo {author} {\bibfnamefont {G.~J.}\ \bibnamefont
  {Fleer}}, \ and\ \bibinfo {author} {\bibfnamefont {E.~B.}\ \bibnamefont
  {Zhulina}},\ }\bibfield  {title} {\enquote {\bibinfo {title} {Charged
  polymeric brushes: Structure and scaling relations},}\ }\href {\doibase
  10.1021/ma00090a018} {\bibfield  {journal} {\bibinfo  {journal}
  {Macromolecules}\ }\textbf {\bibinfo {volume} {27}},\ \bibinfo {pages}
  {3249--3261} (\bibinfo {year} {1994})}\BibitemShut {NoStop}%
\bibitem [{\citenamefont {Zappone}\ \emph {et~al.}(2007)\citenamefont
  {Zappone}, \citenamefont {Ruths}, \citenamefont {Greene}, \citenamefont
  {Jay},\ and\ \citenamefont {Israelachvili}}]{2007Israelachvili1693}%
  \BibitemOpen
  \bibfield  {author} {\bibinfo {author} {\bibfnamefont {B.}~\bibnamefont
  {Zappone}}, \bibinfo {author} {\bibfnamefont {M.}~\bibnamefont {Ruths}},
  \bibinfo {author} {\bibfnamefont {G.~W.}\ \bibnamefont {Greene}}, \bibinfo
  {author} {\bibfnamefont {G.~D.}\ \bibnamefont {Jay}}, \ and\ \bibinfo
  {author} {\bibfnamefont {J.~N.}\ \bibnamefont {Israelachvili}},\ }\bibfield
  {title} {\enquote {\bibinfo {title} {Adsorption, lubrication, and wear of
  lubricin on model surfaces: Polymer brush-like behavior of a glycoprotein},}\
  }\href {\doibase 10.1529/biophysj.106.088799} {\bibfield  {journal} {\bibinfo
   {journal} {Biophys. J.}\ }\textbf {\bibinfo {volume} {92}},\ \bibinfo
  {pages} {1693--1708} (\bibinfo {year} {2007})}\BibitemShut {NoStop}%
\bibitem [{\citenamefont {Attili}, \citenamefont {Borisov},\ and\ \citenamefont
  {Richter}(2012)}]{2012Richter1466}%
  \BibitemOpen
  \bibfield  {author} {\bibinfo {author} {\bibfnamefont {S.}~\bibnamefont
  {Attili}}, \bibinfo {author} {\bibfnamefont {O.~V.}\ \bibnamefont {Borisov}},
  \ and\ \bibinfo {author} {\bibfnamefont {R.~P.}\ \bibnamefont {Richter}},\
  }\bibfield  {title} {\enquote {\bibinfo {title} {Films of end-grafted
  hyaluronan are a prototype of a brush of a strongly charged, semiflexible
  polyelectrolyte with intrinsic excluded volume},}\ }\href {\doibase
  10.1021/bm3001759} {\bibfield  {journal} {\bibinfo  {journal}
  {Biomacromolecules}\ }\textbf {\bibinfo {volume} {13}},\ \bibinfo {pages}
  {1466--1477} (\bibinfo {year} {2012})}\BibitemShut {NoStop}%
\bibitem [{\citenamefont {Cohen}\ \emph {et~al.}(2004)\citenamefont {Cohen},
  \citenamefont {Joester}, \citenamefont {Geiger},\ and\ \citenamefont
  {Addadi}}]{2004Addadi1393}%
  \BibitemOpen
  \bibfield  {author} {\bibinfo {author} {\bibfnamefont {M.}~\bibnamefont
  {Cohen}}, \bibinfo {author} {\bibfnamefont {D.}~\bibnamefont {Joester}},
  \bibinfo {author} {\bibfnamefont {B.}~\bibnamefont {Geiger}}, \ and\ \bibinfo
  {author} {\bibfnamefont {L.}~\bibnamefont {Addadi}},\ }\bibfield  {title}
  {\enquote {\bibinfo {title} {Spatial and temporal sequence of events in cell
  adhesion: From molecular recognition to focal adhesion assembly},}\ }\href
  {\doibase 10.1002/cbic.200400162} {\bibfield  {journal} {\bibinfo  {journal}
  {ChemBioChem}\ }\textbf {\bibinfo {volume} {5}},\ \bibinfo {pages}
  {1393--1399} (\bibinfo {year} {2004})}\BibitemShut {NoStop}%
\bibitem [{\citenamefont {Cuylen}\ \emph {et~al.}(2016)\citenamefont {Cuylen},
  \citenamefont {Blaukopf}, \citenamefont {Politi}, \citenamefont
  {M{\"u}ller-Reichert}, \citenamefont {Neumann}, \citenamefont {Poser},
  \citenamefont {Ellenberg}, \citenamefont {Hyman},\ and\ \citenamefont
  {Gerlich}}]{2016Daniel308}%
  \BibitemOpen
  \bibfield  {author} {\bibinfo {author} {\bibfnamefont {S.}~\bibnamefont
  {Cuylen}}, \bibinfo {author} {\bibfnamefont {C.}~\bibnamefont {Blaukopf}},
  \bibinfo {author} {\bibfnamefont {A.~Z.}\ \bibnamefont {Politi}}, \bibinfo
  {author} {\bibfnamefont {T.}~\bibnamefont {M{\"u}ller-Reichert}}, \bibinfo
  {author} {\bibfnamefont {B.}~\bibnamefont {Neumann}}, \bibinfo {author}
  {\bibfnamefont {I.}~\bibnamefont {Poser}}, \bibinfo {author} {\bibfnamefont
  {J.}~\bibnamefont {Ellenberg}}, \bibinfo {author} {\bibfnamefont {A.~A.}\
  \bibnamefont {Hyman}}, \ and\ \bibinfo {author} {\bibfnamefont {D.~W.}\
  \bibnamefont {Gerlich}},\ }\bibfield  {title} {\enquote {\bibinfo {title}
  {Ki-67 acts as a biological surfactant to disperse mitotic chromosomes},}\
  }\href {\doibase 10.1038/nature18610} {\bibfield  {journal} {\bibinfo
  {journal} {Nature}\ }\textbf {\bibinfo {volume} {535}},\ \bibinfo {pages}
  {308--312} (\bibinfo {year} {2016})}\BibitemShut {NoStop}%
\bibitem [{\citenamefont {Lai}\ and\ \citenamefont
  {Binder}(1992)}]{1992Binder586}%
  \BibitemOpen
  \bibfield  {author} {\bibinfo {author} {\bibfnamefont {P.}~\bibnamefont
  {Lai}}\ and\ \bibinfo {author} {\bibfnamefont {K.}~\bibnamefont {Binder}},\
  }\bibfield  {title} {\enquote {\bibinfo {title} {Structure and dynamics of
  polymer brushes near the {$\theta$} point: A monte carlo simulation},}\
  }\href {\doibase 10.1063/1.463554} {\bibfield  {journal} {\bibinfo  {journal}
  {J. Chem. Phys.}\ }\textbf {\bibinfo {volume} {97}},\ \bibinfo {pages}
  {586--595} (\bibinfo {year} {1992})}\BibitemShut {NoStop}%
\bibitem [{\citenamefont {Grest}\ and\ \citenamefont
  {Murat}(1993)}]{1993Murat3108}%
  \BibitemOpen
  \bibfield  {author} {\bibinfo {author} {\bibfnamefont {G.~S.}\ \bibnamefont
  {Grest}}\ and\ \bibinfo {author} {\bibfnamefont {M.}~\bibnamefont {Murat}},\
  }\bibfield  {title} {\enquote {\bibinfo {title} {Structure of grafted
  polymeric brushes in solvents of varying quality: a molecular dynamics
  study},}\ }\href {\doibase 10.1021/ma00064a019} {\bibfield  {journal}
  {\bibinfo  {journal} {Macromolecules}\ }\textbf {\bibinfo {volume} {26}},\
  \bibinfo {pages} {3108--3117} (\bibinfo {year} {1993})}\BibitemShut {NoStop}%
\bibitem [{\citenamefont {{D.R.M. Williams}}(1993)}]{1993Williams1313}%
  \BibitemOpen
  \bibfield  {author} {\bibinfo {author} {\bibnamefont {{D.R.M. Williams}}},\
  }\bibfield  {title} {\enquote {\bibinfo {title} {Grafted polymers in bad
  solvents: octopus surface micelles},}\ }\href {\doibase 10.1051/jp2:1993202}
  {\bibfield  {journal} {\bibinfo  {journal} {J. Phys. II France}\ }\textbf
  {\bibinfo {volume} {3}},\ \bibinfo {pages} {1313--1318} (\bibinfo {year}
  {1993})}\BibitemShut {NoStop}%
\bibitem [{\citenamefont {Zhulina}, \citenamefont {Singh},\ and\ \citenamefont
  {Balazs}(1998)}]{1998Zhulina1175}%
  \BibitemOpen
  \bibfield  {author} {\bibinfo {author} {\bibfnamefont {E.}~\bibnamefont
  {Zhulina}}, \bibinfo {author} {\bibfnamefont {C.}~\bibnamefont {Singh}}, \
  and\ \bibinfo {author} {\bibfnamefont {A.~C.}\ \bibnamefont {Balazs}},\
  }\bibfield  {title} {\enquote {\bibinfo {title} {Behavior of tethered
  polyelectrolytes in poor solvents},}\ }\href {\doibase 10.1063/1.475498}
  {\bibfield  {journal} {\bibinfo  {journal} {J. Chem. Phys.}\ }\textbf
  {\bibinfo {volume} {108}},\ \bibinfo {pages} {1175--1183} (\bibinfo {year}
  {1998})}\BibitemShut {NoStop}%
\bibitem [{\citenamefont {Pattanayek}, \citenamefont {Pham},\ and\
  \citenamefont {Pereira}(2005)}]{2005Pereira214908}%
  \BibitemOpen
  \bibfield  {author} {\bibinfo {author} {\bibfnamefont {S.~K.}\ \bibnamefont
  {Pattanayek}}, \bibinfo {author} {\bibfnamefont {T.~T.}\ \bibnamefont
  {Pham}}, \ and\ \bibinfo {author} {\bibfnamefont {G.~G.}\ \bibnamefont
  {Pereira}},\ }\bibfield  {title} {\enquote {\bibinfo {title} {Morphological
  structures formed by grafted polymers in poor solvents},}\ }\href
  {http://scitation.aip.org/content/aip/journal/jcp/122/21/10.1063/1.1917772}
  {\bibfield  {journal} {\bibinfo  {journal} {J. Chem. Phys.}\ }\textbf
  {\bibinfo {volume} {122}},\ \bibinfo {eid} {214908} (\bibinfo {year}
  {2005})}\BibitemShut {NoStop}%
\bibitem [{\citenamefont {Carrillo}\ and\ \citenamefont
  {Dobrynin}(2009)}]{2009Dobrynin13158}%
  \BibitemOpen
  \bibfield  {author} {\bibinfo {author} {\bibfnamefont {J.-M.~Y.}\
  \bibnamefont {Carrillo}}\ and\ \bibinfo {author} {\bibfnamefont {A.~V.}\
  \bibnamefont {Dobrynin}},\ }\bibfield  {title} {\enquote {\bibinfo {title}
  {Morphologies of planar polyelectrolyte brushes in a poor solvent: Molecular
  dynamics simulations and scaling analysis},}\ }\href {\doibase
  10.1021/la901839j} {\bibfield  {journal} {\bibinfo  {journal} {Langmuir}\
  }\textbf {\bibinfo {volume} {25}},\ \bibinfo {pages} {13158--13168} (\bibinfo
  {year} {2009})}\BibitemShut {NoStop}%
\bibitem [{\citenamefont {Tagliazucchi}, \citenamefont {de~la Cruz},\ and\
  \citenamefont {Szleifer}(2010)}]{2010Szleifer5300}%
  \BibitemOpen
  \bibfield  {author} {\bibinfo {author} {\bibfnamefont {M.}~\bibnamefont
  {Tagliazucchi}}, \bibinfo {author} {\bibfnamefont {M.~O.}\ \bibnamefont
  {de~la Cruz}}, \ and\ \bibinfo {author} {\bibfnamefont {I.}~\bibnamefont
  {Szleifer}},\ }\bibfield  {title} {\enquote {\bibinfo {title}
  {Self-organization of grafted polyelectrolyte layers via the coupling of
  chemical equilibrium and physical interactions},}\ }\href {\doibase
  10.1073/pnas.0913340107} {\bibfield  {journal} {\bibinfo  {journal} {Proc.
  Natl. Acad. Sci. USA}\ }\textbf {\bibinfo {volume} {107}},\ \bibinfo {pages}
  {5300--5305} (\bibinfo {year} {2010})}\BibitemShut {NoStop}%
\bibitem [{\citenamefont {He}, \citenamefont {Merlitz},\ and\ \citenamefont
  {Sommer}(2014)}]{2014Sommer104911}%
  \BibitemOpen
  \bibfield  {author} {\bibinfo {author} {\bibfnamefont {G.-L.}\ \bibnamefont
  {He}}, \bibinfo {author} {\bibfnamefont {H.}~\bibnamefont {Merlitz}}, \ and\
  \bibinfo {author} {\bibfnamefont {J.-U.}\ \bibnamefont {Sommer}},\ }\bibfield
   {title} {\enquote {\bibinfo {title} {Molecular dynamics simulations of
  polyelectrolyte brushes under poor solvent conditions: Origins of bundle
  formation},}\ }\href
  {http://scitation.aip.org/content/aip/journal/jcp/140/10/10.1063/1.4867466}
  {\bibfield  {journal} {\bibinfo  {journal} {J. Chem. Phys.}\ }\textbf
  {\bibinfo {volume} {140}},\ \bibinfo {eid} {104911} (\bibinfo {year}
  {2014})}\BibitemShut {NoStop}%
\bibitem [{\citenamefont {Lappala}, \citenamefont {Mendiratta},\ and\
  \citenamefont {Terentjev}(2015)}]{2016Terentjev1894}%
  \BibitemOpen
  \bibfield  {author} {\bibinfo {author} {\bibfnamefont {A.}~\bibnamefont
  {Lappala}}, \bibinfo {author} {\bibfnamefont {S.}~\bibnamefont {Mendiratta}},
  \ and\ \bibinfo {author} {\bibfnamefont {E.~M.}\ \bibnamefont {Terentjev}},\
  }\bibfield  {title} {\enquote {\bibinfo {title} {Arrested spinodal
  decomposition in polymer brush collapsing in poor solvent},}\ }\href
  {\doibase 10.1021/ma501985r} {\bibfield  {journal} {\bibinfo  {journal}
  {Macromolecules}\ }\textbf {\bibinfo {volume} {48}},\ \bibinfo {pages}
  {1894--1900} (\bibinfo {year} {2015})}\BibitemShut {NoStop}%
\bibitem [{\citenamefont {Brettmann}\ \emph {et~al.}(2016)\citenamefont
  {Brettmann}, \citenamefont {Laugel}, \citenamefont {Hoffmann}, \citenamefont
  {Pincus},\ and\ \citenamefont {Tirrell}}]{2016Tirrell284}%
  \BibitemOpen
  \bibfield  {author} {\bibinfo {author} {\bibfnamefont {B.~K.}\ \bibnamefont
  {Brettmann}}, \bibinfo {author} {\bibfnamefont {N.}~\bibnamefont {Laugel}},
  \bibinfo {author} {\bibfnamefont {N.}~\bibnamefont {Hoffmann}}, \bibinfo
  {author} {\bibfnamefont {P.}~\bibnamefont {Pincus}}, \ and\ \bibinfo {author}
  {\bibfnamefont {M.}~\bibnamefont {Tirrell}},\ }\bibfield  {title} {\enquote
  {\bibinfo {title} {Bridging contributions to polyelectrolyte brush collapse
  in multivalent salt solutions},}\ }\href {\doibase 10.1002/pola.27959}
  {\bibfield  {journal} {\bibinfo  {journal} {J. Polym. Sci., Part A: Polym.
  Chem.}\ }\textbf {\bibinfo {volume} {54}},\ \bibinfo {pages} {284--291}
  (\bibinfo {year} {2016})}\BibitemShut {NoStop}%
\bibitem [{\citenamefont {Liu}, \citenamefont {Pincus},\ and\ \citenamefont
  {Hyeon}(2017)}]{2017Hyeon1579}%
  \BibitemOpen
  \bibfield  {author} {\bibinfo {author} {\bibfnamefont {L.}~\bibnamefont
  {Liu}}, \bibinfo {author} {\bibfnamefont {P.~A.}\ \bibnamefont {Pincus}}, \
  and\ \bibinfo {author} {\bibfnamefont {C.}~\bibnamefont {Hyeon}},\ }\bibfield
   {title} {\enquote {\bibinfo {title} {{Heterogeneous Morphology and Dynamics
  of Polyelectrolyte Brush Condensates in Trivalent Counterion Solution}},}\
  }\href {\doibase 10.1021/acs.macromol.6b02685} {\bibfield  {journal}
  {\bibinfo  {journal} {Macromolecules}\ }\textbf {\bibinfo {volume} {50}},\
  \bibinfo {pages} {1579--1588} (\bibinfo {year} {2017})}\BibitemShut {NoStop}%
\bibitem [{\citenamefont {Jackson}\ \emph {et~al.}(2017)\citenamefont
  {Jackson}, \citenamefont {Brettmann}, \citenamefont {Vishwanath},
  \citenamefont {Tirrell},\ and\ \citenamefont {de~Pablo}}]{2017dePablo155}%
  \BibitemOpen
  \bibfield  {author} {\bibinfo {author} {\bibfnamefont {N.~E.}\ \bibnamefont
  {Jackson}}, \bibinfo {author} {\bibfnamefont {B.~K.}\ \bibnamefont
  {Brettmann}}, \bibinfo {author} {\bibfnamefont {V.}~\bibnamefont
  {Vishwanath}}, \bibinfo {author} {\bibfnamefont {M.}~\bibnamefont {Tirrell}},
  \ and\ \bibinfo {author} {\bibfnamefont {J.~J.}\ \bibnamefont {de~Pablo}},\
  }\bibfield  {title} {\enquote {\bibinfo {title} {Comparing solvophobic and
  multivalent induced collapse in polyelectrolyte brushes},}\ }\href {\doibase
  10.1021/acsmacrolett.6b00837} {\bibfield  {journal} {\bibinfo  {journal} {ACS
  Macro Letters}\ }\textbf {\bibinfo {volume} {6}},\ \bibinfo {pages}
  {155--160} (\bibinfo {year} {2017})}\BibitemShut {NoStop}%
\bibitem [{\citenamefont {Yu}\ \emph {et~al.}(2017)\citenamefont {Yu},
  \citenamefont {Jackson}, \citenamefont {Xu}, \citenamefont {Brettmann},
  \citenamefont {Ruths}, \citenamefont {de~Pablo},\ and\ \citenamefont
  {Tirrell}}]{2017Tirrell1497}%
  \BibitemOpen
  \bibfield  {author} {\bibinfo {author} {\bibfnamefont {J.}~\bibnamefont
  {Yu}}, \bibinfo {author} {\bibfnamefont {N.~E.}\ \bibnamefont {Jackson}},
  \bibinfo {author} {\bibfnamefont {X.}~\bibnamefont {Xu}}, \bibinfo {author}
  {\bibfnamefont {B.~K.}\ \bibnamefont {Brettmann}}, \bibinfo {author}
  {\bibfnamefont {M.}~\bibnamefont {Ruths}}, \bibinfo {author} {\bibfnamefont
  {J.~J.}\ \bibnamefont {de~Pablo}}, \ and\ \bibinfo {author} {\bibfnamefont
  {M.}~\bibnamefont {Tirrell}},\ }\bibfield  {title} {\enquote {\bibinfo
  {title} {Multivalent ions induce lateral structural inhomogeneities in
  polyelectrolyte brushes},}\ }\href {\doibase 10.1126/sciadv.aao1497}
  {\bibfield  {journal} {\bibinfo  {journal} {Science Advances}\ }\textbf
  {\bibinfo {volume} {3}},\ \bibinfo {pages} {eaao1497} (\bibinfo {year}
  {2017})}\BibitemShut {NoStop}%
\bibitem [{\citenamefont {Bracha}\ and\ \citenamefont
  {Bar-Ziv}(2014)}]{2014BarZiv4945}%
  \BibitemOpen
  \bibfield  {author} {\bibinfo {author} {\bibfnamefont {D.}~\bibnamefont
  {Bracha}}\ and\ \bibinfo {author} {\bibfnamefont {R.~H.}\ \bibnamefont
  {Bar-Ziv}},\ }\bibfield  {title} {\enquote {\bibinfo {title} {Dendritic and
  nanowire assemblies of condensed {DNA} polymer brushes},}\ }\href {\doibase
  10.1021/ja410960w} {\bibfield  {journal} {\bibinfo  {journal} {J. Am. Chem.
  Soc.}\ }\textbf {\bibinfo {volume} {136}},\ \bibinfo {pages} {4945--4953}
  (\bibinfo {year} {2014})}\BibitemShut {NoStop}%
\bibitem [{\citenamefont {Csajka}\ and\ \citenamefont
  {Seidel}(2000)}]{2000Seidel2728}%
  \BibitemOpen
  \bibfield  {author} {\bibinfo {author} {\bibfnamefont {F.~S.}\ \bibnamefont
  {Csajka}}\ and\ \bibinfo {author} {\bibfnamefont {C.}~\bibnamefont
  {Seidel}},\ }\bibfield  {title} {\enquote {\bibinfo {title} {Strongly charged
  polyelectrolyte brushes: A molecular dynamics study},}\ }\href {\doibase
  10.1021/ma990096l} {\bibfield  {journal} {\bibinfo  {journal}
  {Macromolecules}\ }\textbf {\bibinfo {volume} {33}},\ \bibinfo {pages}
  {2728--2739} (\bibinfo {year} {2000})}\BibitemShut {NoStop}%
\bibitem [{\citenamefont {Crozier}\ and\ \citenamefont
  {Stevens}(2003)}]{2003Stevens3855}%
  \BibitemOpen
  \bibfield  {author} {\bibinfo {author} {\bibfnamefont {P.~S.}\ \bibnamefont
  {Crozier}}\ and\ \bibinfo {author} {\bibfnamefont {M.~J.}\ \bibnamefont
  {Stevens}},\ }\bibfield  {title} {\enquote {\bibinfo {title} {Simulations of
  single grafted polyelectrolyte chains: {ssDNA} and {dsDNA}},}\ }\href
  {\doibase http://dx.doi.org/10.1063/1.1540098} {\bibfield  {journal}
  {\bibinfo  {journal} {J. Chem. Phys.}\ }\textbf {\bibinfo {volume} {118}},\
  \bibinfo {pages} {3855--3860} (\bibinfo {year} {2003})}\BibitemShut {NoStop}%
\bibitem [{\citenamefont {Guptha}\ and\ \citenamefont
  {Hsiao}(2014)}]{2014Hsiao2900}%
  \BibitemOpen
  \bibfield  {author} {\bibinfo {author} {\bibfnamefont {V.~S.}\ \bibnamefont
  {Guptha}}\ and\ \bibinfo {author} {\bibfnamefont {P.-Y.}\ \bibnamefont
  {Hsiao}},\ }\bibfield  {title} {\enquote {\bibinfo {title} {Polyelectrolyte
  brushes in monovalent and multivalent salt solutions},}\ }\href {\doibase
  http://dx.doi.org/10.1016/j.polymer.2014.04.035} {\bibfield  {journal}
  {\bibinfo  {journal} {Polymer}\ }\textbf {\bibinfo {volume} {55}},\ \bibinfo
  {pages} {2900 -- 2912} (\bibinfo {year} {2014})}\BibitemShut {NoStop}%
\bibitem [{\citenamefont {Benetatos}, \citenamefont {Terentjev},\ and\
  \citenamefont {Zippelius}(2013)}]{2013Zippelius042601}%
  \BibitemOpen
  \bibfield  {author} {\bibinfo {author} {\bibfnamefont {P.}~\bibnamefont
  {Benetatos}}, \bibinfo {author} {\bibfnamefont {E.~M.}\ \bibnamefont
  {Terentjev}}, \ and\ \bibinfo {author} {\bibfnamefont {A.}~\bibnamefont
  {Zippelius}},\ }\bibfield  {title} {\enquote {\bibinfo {title} {Bundling in
  brushes of directed and semiflexible polymers},}\ }\href {\doibase
  10.1103/PhysRevE.88.042601} {\bibfield  {journal} {\bibinfo  {journal} {Phys.
  Rev. E}\ }\textbf {\bibinfo {volume} {88}},\ \bibinfo {pages} {042601}
  (\bibinfo {year} {2013})}\BibitemShut {NoStop}%
\bibitem [{\citenamefont {Sander}(1986)}]{1986Sander789}%
  \BibitemOpen
  \bibfield  {author} {\bibinfo {author} {\bibfnamefont {L.~M.}\ \bibnamefont
  {Sander}},\ }\bibfield  {title} {\enquote {\bibinfo {title} {Fractal growth
  processes},}\ }\href {\doibase 10.1038/322789a0} {\bibfield  {journal}
  {\bibinfo  {journal} {Nature}\ }\textbf {\bibinfo {volume} {322}},\ \bibinfo
  {pages} {789--793} (\bibinfo {year} {1986})}\BibitemShut {NoStop}%
\bibitem [{\citenamefont {Nicol{\'a}s-Carlock}, \citenamefont
  {Carrillo-Estrada},\ and\ \citenamefont {Dossetti}(2016)}]{2016Dossetti}%
  \BibitemOpen
  \bibfield  {author} {\bibinfo {author} {\bibfnamefont {J.~R.}\ \bibnamefont
  {Nicol{\'a}s-Carlock}}, \bibinfo {author} {\bibfnamefont {J.~L.}\
  \bibnamefont {Carrillo-Estrada}}, \ and\ \bibinfo {author} {\bibfnamefont
  {V.}~\bibnamefont {Dossetti}},\ }\bibfield  {title} {\enquote {\bibinfo
  {title} {Fractality {\`a} la carte: a general particle aggregation model},}\
  }\href {\doibase 10.1038/srep19505} {\bibfield  {journal} {\bibinfo
  {journal} {Sci. Rep.}\ }\textbf {\bibinfo {volume} {6}},\ \bibinfo {pages}
  {19505} (\bibinfo {year} {2016})}\BibitemShut {NoStop}%
\bibitem [{\citenamefont {Nicol{\'a}s-Carlock}, \citenamefont
  {Carrillo-Estrada},\ and\ \citenamefont {Dossetti}(2017)}]{2017Dossetti3523}%
  \BibitemOpen
  \bibfield  {author} {\bibinfo {author} {\bibfnamefont {J.~R.}\ \bibnamefont
  {Nicol{\'a}s-Carlock}}, \bibinfo {author} {\bibfnamefont {J.~L.}\
  \bibnamefont {Carrillo-Estrada}}, \ and\ \bibinfo {author} {\bibfnamefont
  {V.}~\bibnamefont {Dossetti}},\ }\bibfield  {title} {\enquote {\bibinfo
  {title} {Universal fractality of morphological transitions in stochastic
  growth processes},}\ }\href {\doibase 10.1038/s41598-017-03491-5} {\bibfield
  {journal} {\bibinfo  {journal} {Sci. Rep.}\ }\textbf {\bibinfo {volume}
  {7}},\ \bibinfo {pages} {3523} (\bibinfo {year} {2017})}\BibitemShut
  {NoStop}%
\bibitem [{\citenamefont {Frenkel}\ and\ \citenamefont
  {Eppenga}(1985)}]{1986Eppenga1776}%
  \BibitemOpen
  \bibfield  {author} {\bibinfo {author} {\bibfnamefont {D.}~\bibnamefont
  {Frenkel}}\ and\ \bibinfo {author} {\bibfnamefont {R.}~\bibnamefont
  {Eppenga}},\ }\bibfield  {title} {\enquote {\bibinfo {title} {Evidence for
  algebraic orientational order in a two-dimensional hard-core nematic},}\
  }\href {\doibase 10.1103/PhysRevA.31.1776} {\bibfield  {journal} {\bibinfo
  {journal} {Phys. Rev. A}\ }\textbf {\bibinfo {volume} {31}},\ \bibinfo
  {pages} {1776--1787} (\bibinfo {year} {1985})}\BibitemShut {NoStop}%
\bibitem [{\citenamefont {Bates}\ and\ \citenamefont
  {Frenkel}(2000)}]{2000Frenkel10034}%
  \BibitemOpen
  \bibfield  {author} {\bibinfo {author} {\bibfnamefont {M.~A.}\ \bibnamefont
  {Bates}}\ and\ \bibinfo {author} {\bibfnamefont {D.}~\bibnamefont
  {Frenkel}},\ }\bibfield  {title} {\enquote {\bibinfo {title} {Phase behavior
  of two-dimensional hard rod fluids},}\ }\href {\doibase 10.1063/1.481637}
  {\bibfield  {journal} {\bibinfo  {journal} {J. Chem. Phys.}\ }\textbf
  {\bibinfo {volume} {112}},\ \bibinfo {pages} {10034--10041} (\bibinfo {year}
  {2000})}\BibitemShut {NoStop}%
\bibitem [{\citenamefont {Ha}\ and\ \citenamefont {Liu}(1999)}]{1999Liu624}%
  \BibitemOpen
  \bibfield  {author} {\bibinfo {author} {\bibfnamefont {B.~Y.}\ \bibnamefont
  {Ha}}\ and\ \bibinfo {author} {\bibfnamefont {A.~J.}\ \bibnamefont {Liu}},\
  }\bibfield  {title} {\enquote {\bibinfo {title} {Kinetics of bundle growth in
  {DNA} condensation},}\ }\href {\doibase 10.1209/epl/i1999-00311-6} {\bibfield
   {journal} {\bibinfo  {journal} {EPL (Europhysics Letters)}\ }\textbf
  {\bibinfo {volume} {46}},\ \bibinfo {pages} {624--630} (\bibinfo {year}
  {1999})}\BibitemShut {NoStop}%
\bibitem [{\citenamefont {Witten}\ and\ \citenamefont
  {Sander}(1981)}]{1981Sander1400}%
  \BibitemOpen
  \bibfield  {author} {\bibinfo {author} {\bibfnamefont {T.~A.}\ \bibnamefont
  {Witten}}\ and\ \bibinfo {author} {\bibfnamefont {L.~M.}\ \bibnamefont
  {Sander}},\ }\bibfield  {title} {\enquote {\bibinfo {title}
  {Diffusion-limited aggregation, a kinetic critical phenomenon},}\ }\href
  {\doibase 10.1103/PhysRevLett.47.1400} {\bibfield  {journal} {\bibinfo
  {journal} {Phys. Rev. Lett.}\ }\textbf {\bibinfo {volume} {47}},\ \bibinfo
  {pages} {1400--1403} (\bibinfo {year} {1981})}\BibitemShut {NoStop}%
\bibitem [{\citenamefont {Tokuyama}\ and\ \citenamefont
  {Kawasaki}(1984)}]{1984Kawasaki337}%
  \BibitemOpen
  \bibfield  {author} {\bibinfo {author} {\bibfnamefont {M.}~\bibnamefont
  {Tokuyama}}\ and\ \bibinfo {author} {\bibfnamefont {K.}~\bibnamefont
  {Kawasaki}},\ }\bibfield  {title} {\enquote {\bibinfo {title} {Fractal
  dimensions for diffusion-limited aggregation},}\ }\href {\doibase
  https://doi.org/10.1016/0375-9601(84)91083-1} {\bibfield  {journal} {\bibinfo
   {journal} {Phys. Lett. A}\ }\textbf {\bibinfo {volume} {100}},\ \bibinfo
  {pages} {337 -- 340} (\bibinfo {year} {1984})}\BibitemShut {NoStop}%
\bibitem [{\citenamefont {Matsushita}\ \emph {et~al.}(1986)\citenamefont
  {Matsushita}, \citenamefont {Honda}, \citenamefont {Toyoki}, \citenamefont
  {Hayakawa},\ and\ \citenamefont {Kondo}}]{1986Kondo2618}%
  \BibitemOpen
  \bibfield  {author} {\bibinfo {author} {\bibfnamefont {M.}~\bibnamefont
  {Matsushita}}, \bibinfo {author} {\bibfnamefont {K.}~\bibnamefont {Honda}},
  \bibinfo {author} {\bibfnamefont {H.}~\bibnamefont {Toyoki}}, \bibinfo
  {author} {\bibfnamefont {Y.}~\bibnamefont {Hayakawa}}, \ and\ \bibinfo
  {author} {\bibfnamefont {H.}~\bibnamefont {Kondo}},\ }\bibfield  {title}
  {\enquote {\bibinfo {title} {Generalization and the fractal dimensionality of
  diffusion-limited aggregation},}\ }\href {\doibase 10.1143/JPSJ.55.2618}
  {\bibfield  {journal} {\bibinfo  {journal} {J. Phys. Soc. Japan}\ }\textbf
  {\bibinfo {volume} {55}},\ \bibinfo {pages} {2618--2626} (\bibinfo {year}
  {1986})}\BibitemShut {NoStop}%
\bibitem [{\citenamefont {Mima}\ \emph {et~al.}(2017)\citenamefont {Mima},
  \citenamefont {Kinjo}, \citenamefont {Yamakawa},\ and\ \citenamefont
  {Asahi}}]{2017Asahi5991}%
  \BibitemOpen
  \bibfield  {author} {\bibinfo {author} {\bibfnamefont {T.}~\bibnamefont
  {Mima}}, \bibinfo {author} {\bibfnamefont {T.}~\bibnamefont {Kinjo}},
  \bibinfo {author} {\bibfnamefont {S.}~\bibnamefont {Yamakawa}}, \ and\
  \bibinfo {author} {\bibfnamefont {R.}~\bibnamefont {Asahi}},\ }\bibfield
  {title} {\enquote {\bibinfo {title} {Study of the conformation of
  polyelectrolyte aggregates using coarse-grained molecular dynamics
  simulations},}\ }\href {\doibase 10.1039/C7SM01196B} {\bibfield  {journal}
  {\bibinfo  {journal} {Soft Matter}\ }\textbf {\bibinfo {volume} {13}},\
  \bibinfo {pages} {5991--5999} (\bibinfo {year} {2017})}\BibitemShut {NoStop}%
\bibitem [{\citenamefont {Honeycutt}\ and\ \citenamefont
  {Thirumalai}(1992)}]{1992Thirumalai695}%
  \BibitemOpen
  \bibfield  {author} {\bibinfo {author} {\bibfnamefont {J.~D.}\ \bibnamefont
  {Honeycutt}}\ and\ \bibinfo {author} {\bibfnamefont {D.}~\bibnamefont
  {Thirumalai}},\ }\bibfield  {title} {\enquote {\bibinfo {title} {The nature
  of folded states of globular proteins},}\ }\href {\doibase
  10.1002/bip.360320610} {\bibfield  {journal} {\bibinfo  {journal}
  {Biopolymers}\ }\textbf {\bibinfo {volume} {32}},\ \bibinfo {pages}
  {695--709} (\bibinfo {year} {1992})}\BibitemShut {NoStop}%
\bibitem [{\citenamefont {Limbach}\ \emph {et~al.}(2006)\citenamefont
  {Limbach}, \citenamefont {Arnold}, \citenamefont {Mann},\ and\ \citenamefont
  {Holm}}]{2006Holm704}%
  \BibitemOpen
  \bibfield  {author} {\bibinfo {author} {\bibfnamefont {H.~J.}\ \bibnamefont
  {Limbach}}, \bibinfo {author} {\bibfnamefont {A.}~\bibnamefont {Arnold}},
  \bibinfo {author} {\bibfnamefont {B.~A.}\ \bibnamefont {Mann}}, \ and\
  \bibinfo {author} {\bibfnamefont {C.}~\bibnamefont {Holm}},\ }\bibfield
  {title} {\enquote {\bibinfo {title} {{ESPResSo} -- an extensible simulation
  package for research on soft matter systems},}\ }\href@noop {} {\bibfield
  {journal} {\bibinfo  {journal} {Comput. Phys. Commun.}\ }\textbf {\bibinfo
  {volume} {174}},\ \bibinfo {pages} {704--727} (\bibinfo {year}
  {2006})}\BibitemShut {NoStop}%
\bibitem [{\citenamefont {Arnold}\ \emph {et~al.}(2013)\citenamefont {Arnold},
  \citenamefont {Lenz}, \citenamefont {Kesselheim}, \citenamefont {Weeber},
  \citenamefont {Fahrenberger}, \citenamefont {Roehm}, \citenamefont
  {Ko\v{s}ovan},\ and\ \citenamefont {Holm}}]{2013Holm1}%
  \BibitemOpen
  \bibfield  {author} {\bibinfo {author} {\bibfnamefont {A.}~\bibnamefont
  {Arnold}}, \bibinfo {author} {\bibfnamefont {O.}~\bibnamefont {Lenz}},
  \bibinfo {author} {\bibfnamefont {S.}~\bibnamefont {Kesselheim}}, \bibinfo
  {author} {\bibfnamefont {R.}~\bibnamefont {Weeber}}, \bibinfo {author}
  {\bibfnamefont {F.}~\bibnamefont {Fahrenberger}}, \bibinfo {author}
  {\bibfnamefont {D.}~\bibnamefont {Roehm}}, \bibinfo {author} {\bibfnamefont
  {P.}~\bibnamefont {Ko\v{s}ovan}}, \ and\ \bibinfo {author} {\bibfnamefont
  {C.}~\bibnamefont {Holm}},\ }\bibfield  {title} {\enquote {\bibinfo {title}
  {{ESPResSo 3.1 --- Molecular Dynamics Software for Coarse-Grained Models}},}\
  }in\ \href@noop {} {\emph {\bibinfo {booktitle} {Meshfree Methods for Partial
  Differential Equations {VI}}}},\ \bibinfo {series} {Lecture Notes in
  Computational Science and Engineering}, Vol.~\bibinfo {volume} {89},\
  \bibinfo {editor} {edited by\ \bibinfo {editor} {\bibfnamefont
  {M.}~\bibnamefont {Griebel}}\ and\ \bibinfo {editor} {\bibfnamefont {M.~A.}\
  \bibnamefont {Schweitzer}}}\ (\bibinfo  {publisher} {Springer},\ \bibinfo
  {year} {2013})\ pp.\ \bibinfo {pages} {1--23}\BibitemShut {NoStop}%
\end{thebibliography}
%

\appendix
\section{Supplementary Figures}
\renewcommand{\thefigure}{S\arabic{figure}}
\setcounter{figure}{0}

\begin{figure*}[h!]
\centering\includegraphics[width=0.8\linewidth]{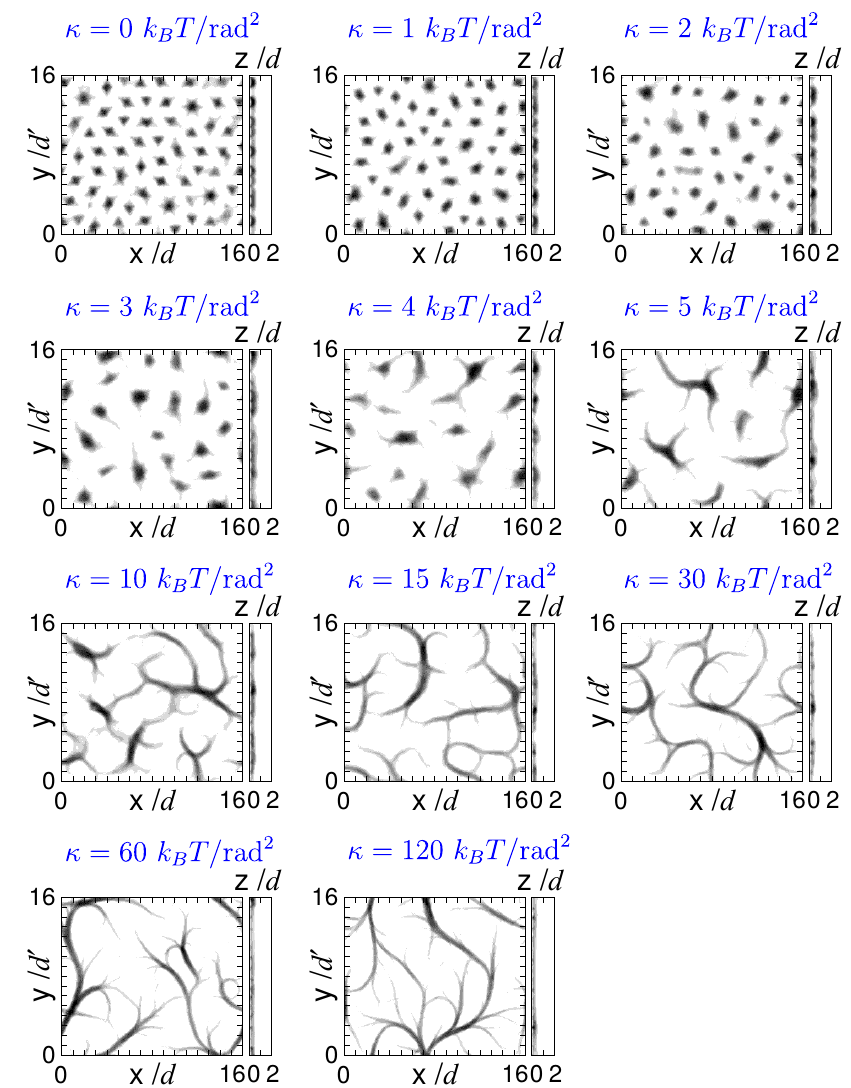}
\caption{
Monomer density heatmap of PE brush averaged over time, 
$\langle \rho(x,y) \rangle$ and $\langle \rho(y,z) \rangle$, with various chain stiffness $\kappa$. 
The length unit $d$ ($=16a$) is the inter-chain distance at the grafting surface, and $d' = \sqrt{3}d/2$ on the $y$-axis.
}
\label{rhoMap}
\end{figure*}

\begin{figure}[h!]
\centering\includegraphics[width=1.\linewidth]{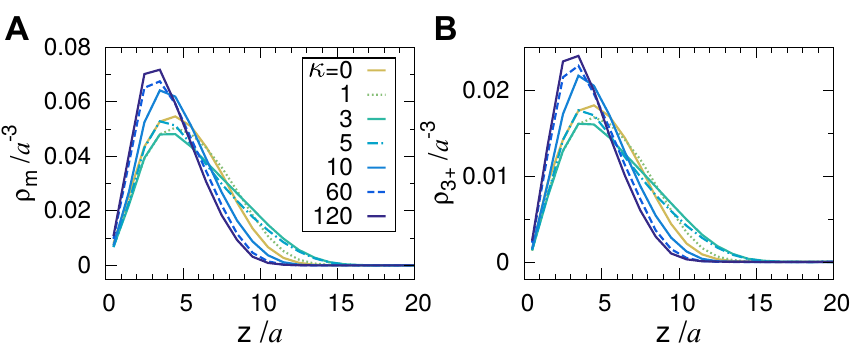}
\caption{
Distribution of (A) chain monomers and (B) trivalent cations (3+) along $z$-axis for the brush chain with different $\kappa$.
}
\label{rhoz}
\end{figure}

\begin{figure}[h!]
\centering\includegraphics[width=1.\linewidth]{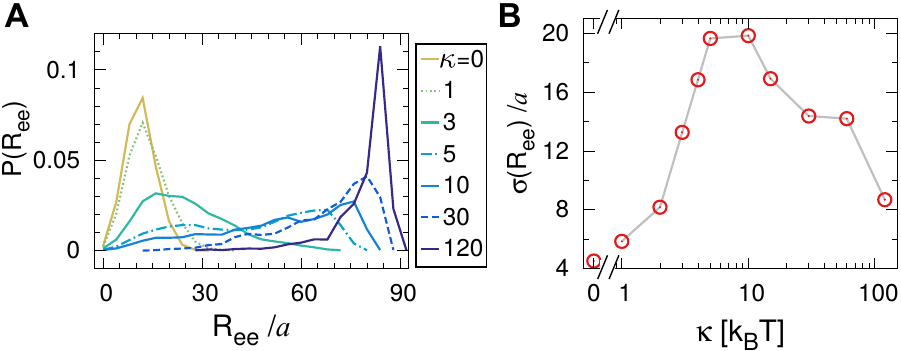}
\caption{
(A) Probability density of the end-to-end distance ($R_{ee}$) of chains in the brush condensate.
(B) Standard deviation of the $R_{ee}$ of brush chains as a function of $\kappa$.
}
\label{chain}
\end{figure}

\begin{figure}[h!]
\centering\includegraphics[width=1.\linewidth]{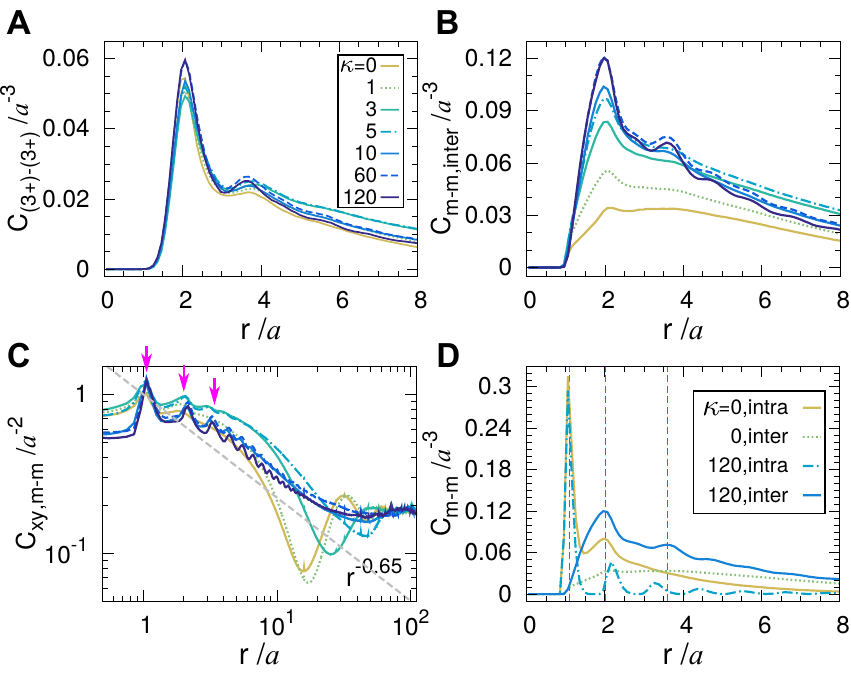}
\caption{
(A) 3D radial distribution function between trivalent ions $C_{(3+)-(3+)}$, 
and (B) between monomers from different chains $C_{m-m,inter}$ at different $\kappa$.
(C) 2D radial distribution function $C_{xy,m-m}$ between chain monomers at different $\kappa$ shown in a log-log scale. 
The gray dashed line indicates a scaling relation of $C(r) \sim r^{-0.65}$. 
(D) 3D radial distribution function $C_{m-m}$ between either intra-chain or inter-chain monomers at $\kappa = 0,120$ $k_{B}T/\text{rad}^2$.
The arrows in (C) and vertical dashed lines in (D) label the length scales $r_{1,2,3}^{\ast}$ at which intermediate scattering function were calculated.
}
\label{cor2d}
\end{figure}

\begin{figure}[h!]
\centering\includegraphics[width=1.\linewidth]{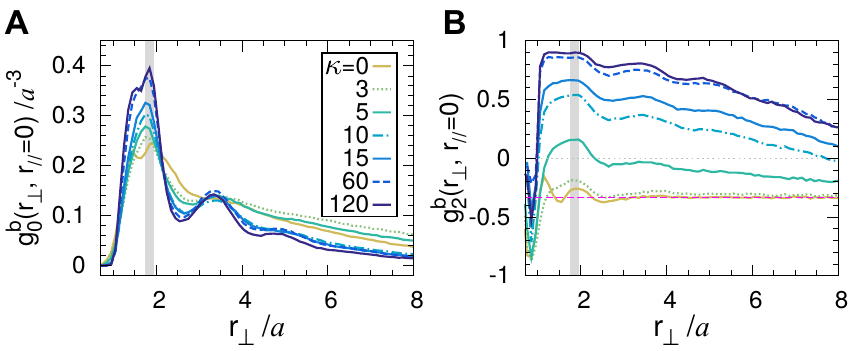}
\caption{
Inter-chain bond ordering. 
(A) $g_{0}^{b}(r_{\perp}, r_{\parallel}=0)$, and (B) $g_{2}^{b}(r_{\perp}, r_{\parallel}=0)$ as a function of $r_{\perp}$ with different $\kappa$, where the gray solid bar 
indicates the positions of the highest peaks. 
}
\label{bdOrderPerp}
\end{figure}

\begin{figure*}[h!]
\centering\includegraphics[width=0.8\linewidth]{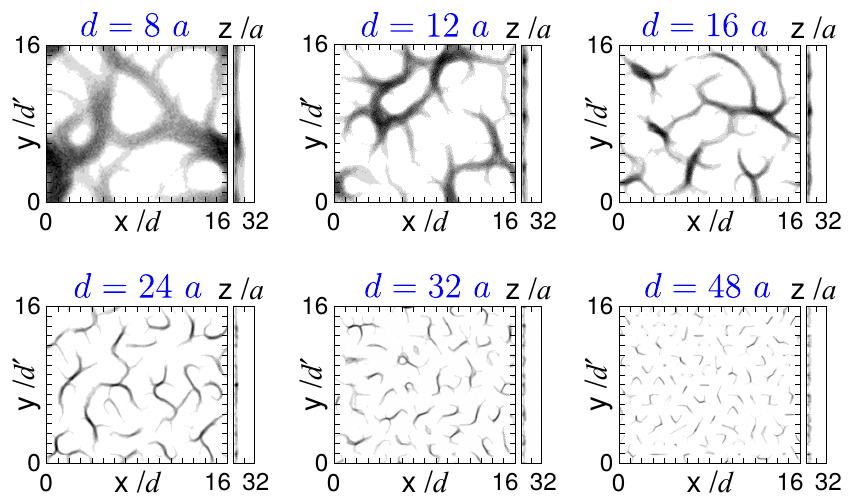}
\caption{
Time-averaged monomer density heatmap of PE brush $\langle \rho(x,y) \rangle$ and $\langle \rho(y,z) \rangle$,
at $\kappa=10$ $k_{B}T/\textrm{rad}^2$, with different inter-chain spacing $d$ on the grafting surface (i.e., with different grafting densities). 
The $x$- and $y$-axes in each plot are scaled with $d$ and $d' = \sqrt{3}d/2$, respectively.
}
\label{rhoMap_g}
\end{figure*}

\begin{figure}[h!]
\centering\includegraphics[width=1.\linewidth]{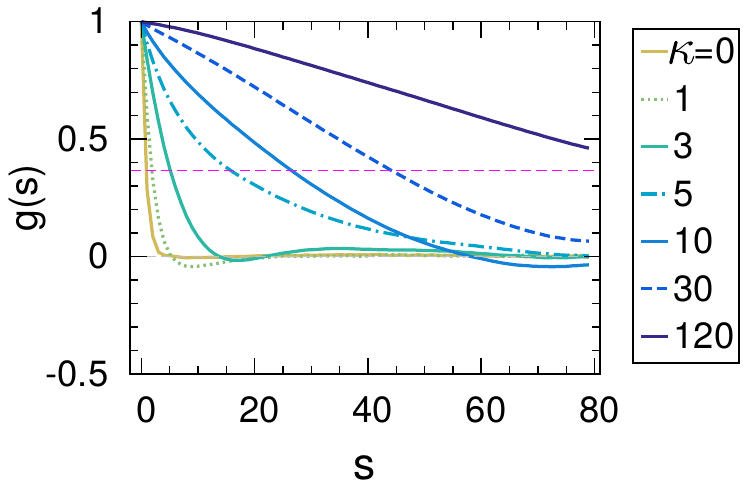}
\caption{
Bond orientation correlation of chains in the brush, 
defined as $g(s) = \sum_{i}^{N-s} \vec{u}_{i} \cdot \vec{u}_{i+s} / (N-s)$, 
with a dashed magenta line indicating $g(l_{p}) = 1/e$.
}
\label{gs0}
\end{figure}

\end{document}